\documentclass[aps,prd,floatfix,showpacs,superscriptaddress,nofootinbib,notitlepage]{revtex4-1}
\usepackage[colorlinks=true, pdfstartview=FitV, linkcolor=blue, citecolor=red, urlcolor=magenta]{hyperref}
\usepackage[utf8]{inputenc}
\usepackage{graphicx,ulem}
\usepackage{latexsym}
\usepackage{amsmath}
\usepackage{amsfonts}
\usepackage{amssymb}
\usepackage{bbm}
\usepackage{verbatim}
\usepackage{appendix}
\usepackage{float}
\usepackage{braket}
\newcommand{\MSbar}{\overline{\mbox{MS}}}

\def\b{\beta}
\def\m{\mu}

\def\n{\nu}
\def\r{\rho}
\def\l{\lambda}

\def\r{\rho}

\def\d{\partial}

\def\vf{\varphi}

\def\dd #1 #2{{\delta #1\over \delta #2}}

\def\ha{\frac{1}{2}}

\newcommand{\pa}{\partial}
\def\n{v}
\def\m{\mu}
\def\n{\nu}

\newcommand{\beq}{\begin{eqnarray}}
\newcommand{\eeq}{\end{eqnarray}}

\newcommand{\kulak}{KU Leuven Campus Kortrijk---Kulak, Department of Physics, Etienne Sabbelaan 53 bus 7657, 8500 Kortrijk, Belgium}
\newcommand{\ughent}{Ghent University, Department of Physics and Astronomy, Krijgslaan 281-S9, 9000 Gent, Belgium}
\newcommand{\uerj}
{Universidade do Estado do Rio de Janeiro,
	Instituto de F\'isica---Departamento de F\'isica Te\'orica---Rua S\~{a}o Francisco Xavier 524,
	20550-013, Maracan\~{a}, Rio de Janeiro, Brasil}

\begin{document}
	\title{  Gauge-invariant spectral description of the $U(1)$ Higgs model from local composite operators
		}

	\author{D.~Dudal }\email{david.dudal@kuleuven.be}\affiliation{\kulak}\affiliation{\ughent}

	\author{D.~M.~van Egmond }\email{duifjemaria@gmail.com}\affiliation{\uerj}
	
	\author{M.~S.~Guimar\~{a}es}\email{msguimaraes@uerj.br}\affiliation{\uerj}
	\author{O.~Holanda }\email{ozorio.neto@uerj.br}\affiliation{\uerj}
	\author{L.~F.~Palhares}\email{leticia.palhares@uerj.br}\affiliation{\uerj}
	\author{G.~Peruzzo}\email{gperuzzofisica@gmail.com}\affiliation{\uerj}
	\author{S.~P.~Sorella}\email{silvio.sorella@gmail.com}\affiliation{\uerj}

	\begin{abstract}
The spectral properties of a set of local gauge-invariant composite operators are investigated in the $U(1)$ Higgs model
		quantized in the 't Hooft $R_{\xi}$ gauge. These operators enable us to give
		a gauge-invariant  description of the spectrum of the  theory,   thereby surpassing certain incommodities when using the standard elementary fields.
		The corresponding  two-point correlation functions are evaluated at one-loop
		order and their spectral functions are obtained explicitly. As expected, the above
		mentioned correlation functions   are independent from the
		gauge parameter $\xi$, while exhibiting positive spectral densities as well
		as   gauge-invariant pole masses corresponding to the massive photon and
		Higgs  physical excitations.
	\end{abstract}
	
	\maketitle
	
	\section{Introduction}
	
	An essential aspect of gauge theories is that all physical observable quantities have to be gauge-invariant \cite{Peskin:1995ev,tHooft:1980xss}. However, in practice, the explicit calculations of the $S$-matrix elements and corresponding cross sections are done by employing the   non-gauge-invariant elementary fields such as the $W$ bosons and the Higgs field of the electroweak theory, giving results in quite accurate agreement with experimental ones. 

The success of making use of the non-gauge-invariant elementary fields can be traced back to the so called Nielsen identities \cite{Nielsen:1975fs,Piguet:1984js,gambino1999fermion,gambino2000nielsen,Grassi:2000dz} which follow from the Slavnov-Taylor identities encoding the BRST symmetry of quantized gauge theories. The Nielsen identities ensure that the pole masses of both transverse gauge bosons and Higgs field propagators do not depend on the gauge parameters entering the gauge fixing condition, a pivotal property shared by the $S$-matrix elements. Nevertheless, as one can easily figure out, the use of the non-gauge-invariant fields has its own limitations which show up in several ways. For example, the analysis of the spectral properties of the elementary two-point correlation functions in terms of the K\"all\'{e}n-Lehmann (  KL) representation is often plagued by an undesired dependence of the spectral densities on the gauge parameters and/or the densities attaining negative values, obscuring their physical interpretation.   Indeed, from e.g.~non-perturbative lattice QCD studies, it is well known that not only direct particle-spectrum related properties are hiding in the spectral functions, but at finite temperature also information on transport properties in the quark-gluon plasma etc., see for instance \cite{Asakawa:2000tr,Asakawa:2003re,Aarts:2007pk,Meyer:2011gj,Rothkopf:2019ipj}.  The spectral functions considered are those of gauge-invariant operators.   Moreover, it is also known that in certain classes of gauges, the Nielsen identities can suffer from fatal infrared singularities \cite{Nielsen:1975fs,Aitchison:1983ns,Andreassen:2014eha}, obscuring what happens with e.g.~the pole mass or effective potential governing the Higgs vacuum expectation value in such gauges. 

A formulation of the properties of the   observable excitations in terms of gauge-invariant variables is thus very welcome. Such an endeavour has been addressed by several authors\footnote{See \cite{Maas:2017wzi,maas2019observable}   for a general review. }   \cite{hooft1980we,hooft2012nonperturbative,Frohlich:1980gj,Frohlich:1981yi}, who have been able to construct, out of the elementary fields, a set of local gauge-invariant composite operators which can  effectively implement a gauge-invariant framework by using the tools of quantum gauge field theories: renormalizability, locality, Lorentz covariance and BRST exact symmetry. 

The aim of this work, which generalizes a previous one \cite{us}  devoted to the study of the analytic properties of the propagators of the non gauge-invariant elementary fields, is that of discussing the features of two local gauge-invariant operators within the framework of the $U(1)$ Abelian Higgs model, whose action is specified by
	\beq
	S_0 = \int d^4x \left\{\frac{1}{4} F_{\m\n}F_{\m\n} + (D_{\m}\vf)^{\dagger} D_{\m}\vf +\frac{\l}{2}\left(\vf^{\dagger}\vf-\frac{v^2}{2}\right)^2\right\}\label{higgsqed},
	\eeq
	where the photon field-strength tensor and the covariant derivative are respectively given by
	\beq
	F_{\m\n}=\pa_{\m}A_{\n}-\pa_{\n}A_{\m},\nonumber\\
	D_{\m}\vf = \pa_{\m}\vf+ieA_{\m}\vf
	\eeq
	and the scalar field may be decomposed to account for the Higgs mechanism as
	\beq
	\vf=\frac{1}{\sqrt{2}}((v+h)+i\r)\label{higgs} \;,
	\eeq
	with $h$ and $\rho$ denoting, respectively, the Higgs and the Goldstone fields, while $v$ is the classical minimum of the Higgs potential of eq.\eqref{higgsqed}, responsible for the photon mass generation in this Higgs model. 	
	The action $S_0$ is left invariant by the following gauge transformations
	\beq
	\delta A_{\m}&=&-\pa_{\m}\omega,\quad\delta \vf = ie\omega\vf,\quad\delta \vf^{\dagger}=-ie\omega\vf^{\dagger},\nonumber\\
	\delta h &=&-e\omega\r,\quad \delta \r =e\omega(v+h) \;,
	\label{pppo}
	\eeq
	where $\omega$ is the gauge parameter. 

Following \cite{hooft1980we,hooft2012nonperturbative},  we shall consider the two local  composite operators $O(x)$ and $V_\mu(x)$ invariant under \eqref{pppo}, given by
	\beq
	O(x) &=& 1/2 (h^2(x) +2 v h(x) + \rho^2(x))=\vf^{\dagger}(x) \vf(x)-\frac{v^2}{2} \;,\nonumber\\
	V_{\mu}(x) &=&  -i \vf^{\dagger}(x) (D_\mu \vf)(x) \;.  \label{op12}
	\eeq
	The relevance of these operators can be understood by using the expansion \eqref{higgs} and retaining the first order terms. For the two-point correlator of the scalar operator one finds (cf.~eq.~\eqref{exp1} for the full expression):
	\beq 
	\braket{{O}(x) {O}(y)}  &\sim&  v^2\langle h(x) h(y) \rangle_{\text{tree level}} + {\cal O}(\hbar) + \braket{{\cal O}\left(h^3;h\rho^2;\rho^4\right)} \;,  \label{mean}
	\eeq
	while the contributions to the vector operator at lowest order in the fields read
	\beq	
	V_{\mu}(x) &\sim & \frac{e v^2}{2} A_\mu(x)  + {\rm total\; derivative}\; +\; { \rm higher \; \; orders} \;.  \label{mean2}
	\eeq
	We see therefore that the gauge-invariant operator ${O}(x)$ is related to the Higgs excitation, while $ V_{\mu}(x) $ is associated with the photon. 

	In the sequel, we shall compute the BRST invariant two-point correlation functions
	\beq
	\langle O(x) O(y) \rangle\;,   \; \; \; \; \; \; \langle V_{\mu}(x) V_{\nu}(y) \rangle \;, \label{cf}
	\eeq
	at one-loop order in the 't Hooft  $R_{\xi}$-gauge and discuss the differences with respect to the corresponding one-loop elementary propagators $\langle h(x) h(y) \rangle$ and $\langle A_\mu(x) A_\nu(y) \rangle$ already evaluated in \cite{us} . 

As expected, both correlation functions of eq.\eqref{cf}
	turn out to be independent from the gauge parameter $\xi$. Moreover, we shall show that the one-loop pole masses of $\langle V_{\mu}(x) V_{\nu}(y) \rangle^T $ and $\langle O(x) O(y) \rangle$ are exactly the same as those of the elementary propagators $\langle A_\mu(x) A_\nu(y) \rangle^{T}$ and
	$\langle h(x) h(y) \rangle$, where $\langle A_\mu(x) A_\nu(y) \rangle^{T}$ stands for the transverse component of $\langle A_\mu(x) A_\nu(y) \rangle$,  {\it i.e.}
	\beq
	\langle A_\mu(x) A_\nu(y) \rangle^{T} = \left(\delta_{\mu \rho} - \frac{\partial_\mu \partial_\rho}{\partial^2}\right)\langle A_\rho(x) A_\nu(y) \rangle\, .
	\eeq
	This important feature makes apparent that the operators $V_{\mu}(x)$ and  $ O(x) $ give a gauge-invariant picture for the photon and Higgs modes. In addition, the correlation functions $\langle V_{\mu}(x) V_{\nu}(y) \rangle^T $ and $\langle O(x) O(y) \rangle$ exhibit a spectral KL representation with positive spectral densities, allowing for a physical interpretation in terms of   observable particles. This property is in sharp contrast with the one-loop spectral density of the elementary non-gauge-invariant Higgs propagator $\langle h(x) h(y) \rangle$, which displays an explicit dependence on the gauge parameter $\xi$ \cite{us}. Moreover, the longitudinal part of the correlator $\langle V_{\mu}(x) V_{\nu}(y) \rangle$ ---which is independently gauge-invariant--- is shown to exhibit the pole mass of the Higgs excitation. This last feature reinforces the consistency of the present description of the physical degrees of freedom of the theory, since the only physically expected elementary excitations are indeed the Higgs and the photon ones.
	Let us also underline that, to our knowledge, this is the first explicit one-loop calculation of the gauge-invariant correlators \eqref{cf} and of their analytical properties   in Higgs-like models. 

This work is organized as follows. In section \ref{1a}, we give a short review of the $U(1)$ Abelian Higgs model and of its quantization in the $R_{\xi}$-gauge. Then, we compute at one-loop order the two-point functions: for the elementary fields in \ref{elm} and for the composite operators in \ref{fmsp}.   We pay attention on how to partially resum contributions to the connected propagator. This is of particular relevance when considering a composite operator propagator, where the standard connection of the $1PI$ self-energy being the inverse connected propagator is lost. In section \ref{2a}, we present  an overview of the techniques employed in \cite{us} to obtain the spectral function up to first order in $\hbar$. In section \ref{IV} we review, for the benefit of the reader,  the results for the spectral functions of the propagators of the elementary fields \cite{us}. In section \ref{2c}, we provide the detailed analysis of the computation of the gauge-invariant correlators \eqref{cf} and compare them with the corresponding correlators of the elementary fields.   We connect the subtracted spectral KL representations with the contact terms that can be added to the action in presence of composite operators. The unitary limit, in which the gauge parameter $\xi$ tends to infinity, is investigated in Section \ref{unitarylimit},   where we also make the connection with the gauge-invariant spectral densities. Section \ref{VI} collects our conclusion and outlook.  The final Appendices contain the derivation of the Feynman rules and of the one-loop diagrams contributing to \eqref{cf}.

	\section{  The gauge-invariant operators $V_{\mu}(x)$ and $O(x)$   in the $U(1)$ Higgs Model\label{II}}
	In this section, we will follow the steps outlined in \cite{Maas:2017wzi,torek2016testing,maas2019observable} to obtain the two-point functions for the composite gauge-invariant operators $(V_{\mu}(x), O(x))$ in the Abelian Higgs model. In \ref{1a} we shall  lay out some of the essential properties of the Abelian Higgs model quantized in the $R_{\xi}$-gauge.  The   cancellation of the gauge parameter $\xi$ will help us to  verify  the explicit gauge  independence of  the correlation functions \eqref{cf}.  In \ref{elm} we will shortly review the expressions of  the two-point functions of the elementary fields, obtained in \cite{us}. In \ref{fmsp} we shall compute the two-point function of the two composite gauge-invariant operators $(V_{\mu}(x), O(x))$.
	\subsection{The Abelian Higgs Model: some essentials \label{1a}}
	We start from the $U(1) $ Abelian Higgs classical action as given in eq.\eqref{higgsqed}. The parameter $v$, corresponding to the   minimum of the classical potential present in the starting action,  gives the vacuum expectation value (vev) of the scalar field  to zeroth order in $\hbar$ , $\langle \vf \rangle_0 =v$.
	As usual, the Higgs mechanism \cite{Higgs:1964pj,Higgs:1964ia,Englert:1964et,Guralnik:1964eu} is implemented by expressing  the scalar field as an expansion around its vev,  namely
	\beq
	\vf=\frac{1}{\sqrt{2}}((v+h)+i\r)\label{higgsd},
	\eeq
	where the real part $h$ is identified as  the Higgs field and $\rho$ is the (unphysical) Goldstone boson, with $\langle \rho \rangle=0$.  Here we choose to expand around the classical value of the vev\footnote{  In principle, a non-perturbative gauge-invariant setup implies that $\langle\vf\rangle = 0$, so that this expansion with $\langle\vf\rangle \ne 0$ is only well-defined in the gauge fixed framework that will be described in the next subsection. As is well-known the gauge fixed description of the Higgs mechanism is a successful approach to perturbative calculations in the continuum as the one pursued in the current work. For a more thorough discussion on gauge invariance and the Higgs mechanism, the reader is referred to \cite{Frohlich:1980gj,Frohlich:1981yi}.}, so that $\langle h \rangle$ is zero at the classical level, but receives loop corrections\footnote{ There is of course an equivalent procedure of fixing $\langle h \rangle$ to zero at all orders, by expanding $\vf$ around the full vev:
		$\vf=\frac{1}{\sqrt{2}}((\langle \vf \rangle+h)+i\r)$. See \cite{us} for details.}. The action \eqref{higgsqed} now becomes
	\beq
	S_0&=&\int d^4 x \,\left\{\frac{1}{4} F_{\m\n}F_{\m\n}+\ha\pa_{\m}h\pa_{\m}h+\ha\pa_{\m}\r\pa_{\m}\r - e\,\r\,\pa_{\m}h\, A_{\m}+e\,(h+v)A_{\m}\pa_{\m}\r\right. \nonumber\\
	&+&\left.\frac{1}{2} e^2 A_{\m}[(h+v)^2 + \r^2]A_{\m}+\frac{1}{8}\l(h^2+2h v +\r^2)^2\right\}\label{fullaction2}
	\eeq
	and we notice that both the gauge field and the Higgs field have acquired the following masses
	\beq
	m^2 = e^2 v^2,\,\, m_{h}^2 = \l v^2.
	\eeq
	With this parametrization, the Higgs coupling $\lambda$ and  the parameter $v$ can be fixed in terms of $m$, $m_h$ and $e$, whose values  will be suitably chosen later on in the text.

	\subsubsection{Gauge fixing}
	Quantization of the theory \eqref{fullaction2} requires a proper gauge fixing. We shall employ the gauge fixing term
	\beq
	S_{gf}=\int d^4x \left\{\frac{1}{2\xi}\left(\partial_{\m}A_{\m}+\xi m \rho\right)^2\right\},
	\label{34}
	\eeq
	known as the 't Hooft or $R_{\xi}$-gauge, which has the pleasant property of cancelling the mixed term $\int d^4x (ev\;A_{\m}\pa_{\m}\r)$ in the expression \eqref{fullaction2}. Of course, \eqref{34} breaks the gauge invariance of the action. As is well known, the latter is replaced by the BRST invariance. In fact, introducing the FP ghost fields $\bar{c},c$ as well as the  auxiliary field $b$, for the BRST transformations we have  
	\beq
	sA_{\m}&=&-\pa_{\m}c,\quad	s c~=~ 0,\quad	s \vf ~=~ iec\vf, \quad s\vf^{\dagger} ~=~ -ie c \vf ^{\dagger},\nonumber\\
	s h &=& -e c \r,\quad	s\r ~=~ e c(v+ h),\quad	s\bar{c}~=~ib,\quad	s b~=~0.
	\label{brst}
	\eeq
	Importantly, the operator $s$ is nilpotent, i.e.~$s^2=0$, allowing to work with the so-called BRST  cohomology \cite{Piguet:1995er}, a useful concept to prove unitarity and renormalizability of the Abelian Higgs model \cite{Becchi:1974md,Becchi:1974xu,kugo1979local},   see also \cite{Kraus:1995jk}.
We can now introduce the gauge fixing in a BRST invariant way via
	\beq
	\mathcal{S}_{gf}&=&s\int d^d x \left\{-i\frac{\xi}{2}\bar{c}b+\bar{c}(\pa_{\m}A_{\m} +\xi m \r)\right\}, \nonumber \\
	&=&\int d^d x \left\{\frac{\xi}{2}b^2+ib\pa_{\mu}A_{\m}+i b\xi m \r+\bar{c}\pa^2c-\xi m^2 \bar{c}c- \xi me\bar{c}hc\right\}.
	\eeq
	Notice that the ghosts $(\bar c,c)$ get a gauge parameter-dependent mass, while interacting directly  with the Higgs field.
	
	The total gauge fixed BRST-invariant action then becomes
	\beq
	S&=& S_0+S_{gf}=\int d^4 x \,\Bigg\{\frac{1}{4} F_{\m\n}F_{\m\n} +\ha\pa_{\m}h\pa_{\m}h+\ha\pa_{\m}\r\pa_{\m}\r - e\,\r\,\pa_{\m}h\, A_{\m}+ e\,h A_{\m}\pa_{\m}\r+ \frac{1}{2} m^2 A_{\mu}A_{\mu} \nonumber\\
	&+& \frac{1}{2} e^2 A_{\m}[h^2 +2vh+ \r^2]A_{\m}+\frac{1}{8}\l(h^2 +\r^2)(h^2 +\r^2+4h v)+\ha m_h^2 h^2+m A_{\mu}\pa_{\mu}\rho+\frac{\xi}{2}b^2+ib \pa_{\m}A_{\m} \nonumber\\
	&+&ib \xi m \rho +\bar{c}(\pa^2)c - m^2 \xi c\bar{c}-m\xi e \bar{c}c h\Bigg\}, \label{fullaction}
	\eeq
	with
	\beq
	s S = 0\;. \label{brstinvact}
	\eeq
	
	In Appendix \ref{FR} we collect the propagators and vertices corresponding to the action \eqref{fullaction} of the Abelian Higgs model in the $R_{\xi}$ gauge. 

Let us end this section by pointing out that the two local operators $(V_{\mu}(x), O(x))$ belong to the cohomology of the BRST operator \cite{Piguet:1995er}, {\it i.e.}
	\beq
	s V_{\mu}(x) & = & 0 \;, \qquad \qquad V_{\mu}(x) \neq s\Delta_\mu(x) \nonumber \\
	s O(x) & = & 0 \;, \qquad \qquad O(x) \neq s \Delta(x) \label{coh} \;,
	\eeq
	for any local quantities $(\Delta_\mu(x), \Delta(x))$.
	\subsection{One-loop propagators for the  elementary fields \label{elm}}
	In \cite{us}, we studied the spectral properties of the one-loop propagators for the photon field $A_{\m}(x)$ and the Higgs field $h(x)$ and evaluated them for $d=4$ through dimensional regularization in the $\MSbar$-scheme. 

\begin{figure}[t]
			\center
			\includegraphics[width=12cm]{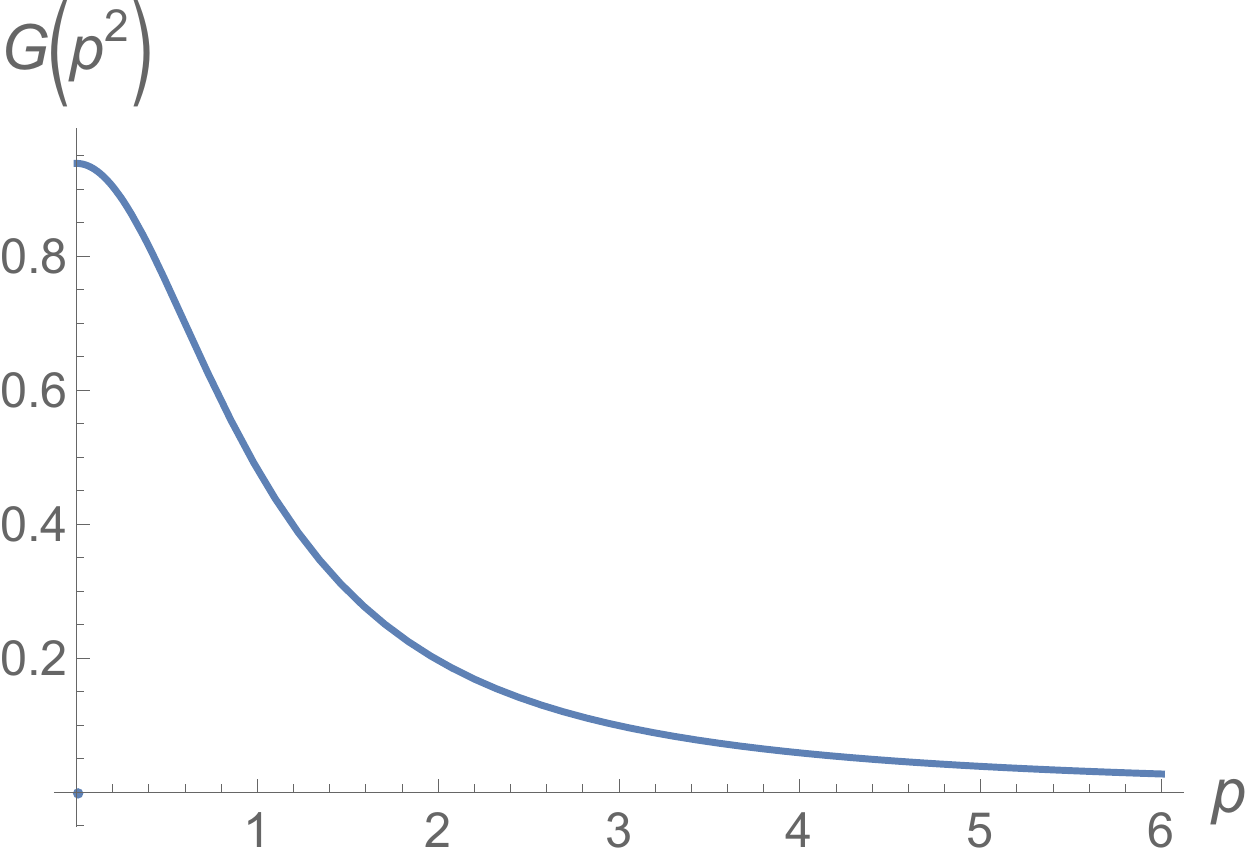}
			\caption{Resummed   photon propagator, with all quantities given in units of appropriate powers of the energy scale $\mu$, for the parameter values $e=1$, $v=1 \, \mu$, $\lambda=\frac{1}{5}$.}
			\label{propAAA}
		\end{figure}

For the photon field, the transverse part of the   connected propagator $G_{\m\n}^{AA}(p^2)$ up to order $\hbar$ is given in  momentum space by
	\beq
	\braket{A_{\mu}(p) A_{\n}(-p)}&=&\frac{1}{p^2+m^2}+\frac{1}{(p^2+m^2)^2}\Pi_{AA}(p^2)+\mathcal{O}(\hbar^2)
	\eeq
	with   the self-energy given by
		\beq
	\Pi_{AA}(p^2)&=& 2\frac{e^2}{(4\pi)^2}\int_{0}^{1} dx \,\,\Bigg\{p^2 x(1-x)+m^2x \nonumber\\
	&+&m_h^2(1-x)(1-\ln\frac{p^2 x(1-x)+m^2x+m_h^2(1-x)}{\m^2})+m_h^2(1-\ln\frac{m_h^2}{\m^2})\nonumber\\
	&+&\frac{m^4}{m_h^2}(1-3 \ln \frac{m^2}{\m^2})+2m^2 \ln \frac{p^2 x(1-x)+m^2x+m_h^2(1-x)}{\m^2}\Bigg\},
	\eeq
	and we can   resum all one-loop self-energy insertions into the connected propagator via 
		\beq
	G^{T}_{AA} (p^2) &=& \frac{1}{ p^2+m^2-\Pi(p^2)},
	\label{dk}
	\eeq
	shown in FIG.~\ref{propAAA}.
	
\begin{figure}[t]
			\center
			\includegraphics[width=12cm]{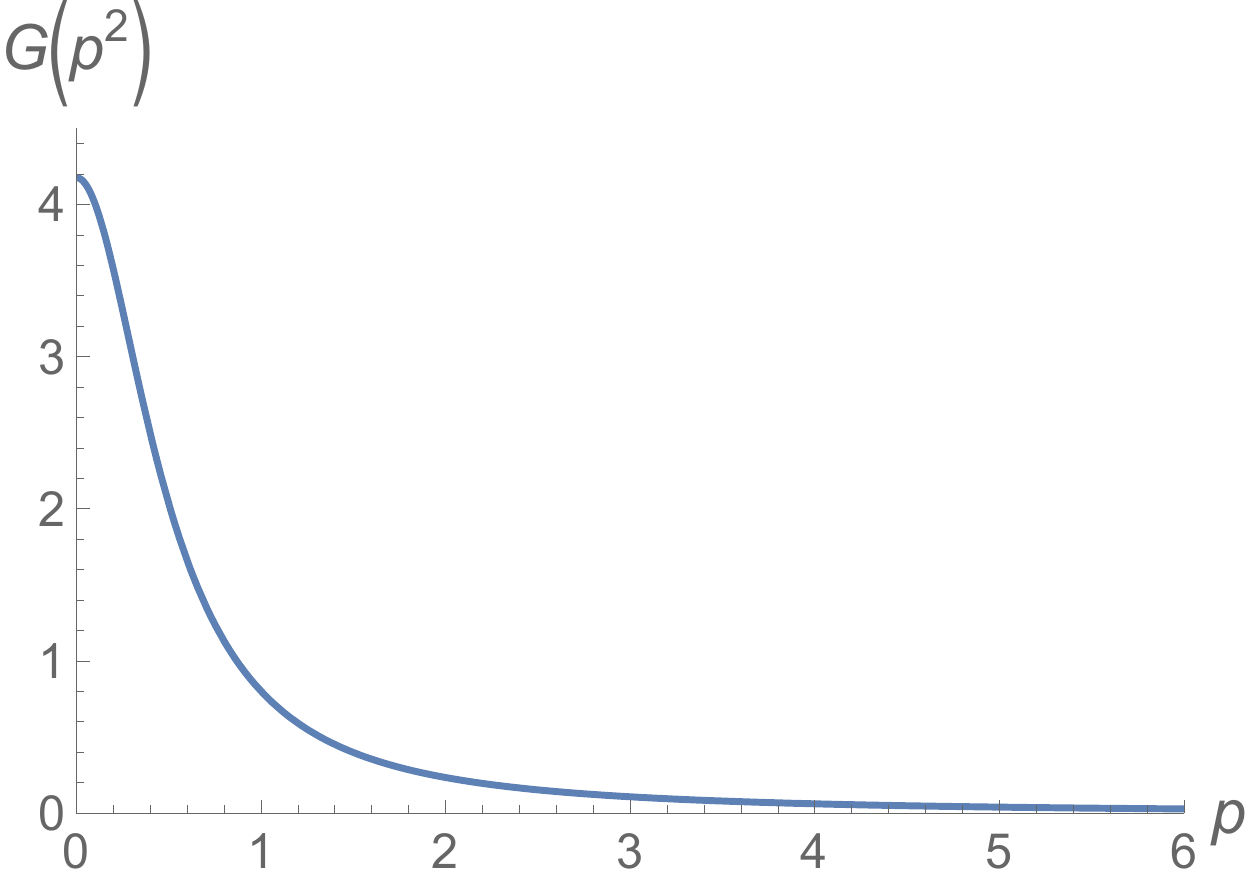}
			\caption{Resummed   Higgs propagator, with all quantities given in units of appropriate powers of the energy scale $\mu$, for the parameter values $e=1$, $v=1 \, \mu$, $\lambda=\frac{1}{5}$.}
			\label{prophhh}
		\end{figure}

	For the Higgs field, we find
	\beq
	\braket{h(p) h(-p)}&=&\frac{1}{p^2+m_h^2}+ \frac{1}{(p^2+m_h^2)^2} \Pi_{hh}(p^2)+\mathcal{O}(\hbar^2)
	\label{kik}
	\eeq
	with
	\beq
	\Pi_{hh}(p^2)&=& \frac{1}{(4\pi)^2}\int_{0}^{1} dx
	\Bigg\{
	e^2\Bigg[p^2(1-\ln\frac{m^2}{\m^2}-2\ln \frac{p^2 x(1-x)+m^2}{\m^2}) \nonumber\\
	&-&\frac{p^4}{2m^2} \ln \frac{p^2 x(1-x)+m^2}{\m^2}-6m^2(1-\ln \frac{m^2}{\m^2}+\ln \frac{p^2 x(1-x)+m^2}{\m^2})\Bigg]\nonumber\\
	&+&\lambda \Big[\frac{1}{2}m_h^2(-6+6\ln \frac{m_h^2}{\m^2}-9\ln \frac{p^2 x(1-x)+m_h^2}{\m^2})\Big]\nonumber\\
	&-&\Bigg[\xi (e^2p^2+\l m^2)(1- \ln \frac{\xi m^2}{\m^2})-(e^ 2\frac{p^4}{2m^2}-\lambda\frac{m_h^2}{2})\ln \frac{p^2 x(1-x)+\xi m^2}{\m^2}\Bigg]
	\Bigg\}.
	\label{dk2}
	\eeq
  Before trying to resum the self-energy insertions again, we notice that this resummation   is tacitly assuming that the second term in \eqref{kik} is much smaller than the first term. Then, we see that   eq.~\eqref{kik} contains terms of the order of $ \frac{p^4}{(p^2+m_h^2)^2} \ln \frac{p^2x(1-x)+m_h^2}{\mu^2}$, which cannot be resummed for big values of $p$. We therefore use the identity
\beq
p^4 &=& (p^2+m_h^2)^2-m_h^4-2p^2m_h^2,
\label{jju1}
\eeq
to rewrite
\beq
  \frac{p^4}{(p^2+m_h^2)^2} \ln \frac{p^2x(1-x)+m_h^2}{\mu^2} &=& \ln \frac{p^2x(1-x)+m_h^2}{\mu^2} -\underline{\frac{(m^4+2p^2m^2)}{(p^2+m^2)^2} \ln \frac{p^2x(1-x)+m_h^2}{\mu^2}}.
\label{jju}
\eeq
The   underlined term in \eqref{jju} can be safely resummed. We thence rewrite
	\beq
	\frac{{\Pi}_{hh}(p^2)}{(p^2+m_h^2)^2}&=& \frac{\hat{\Pi}_{hh}(p^2)}{(p^2+m_h^2)^2}+C_{hh}(p^2),
	\label{jju2}
	\eeq
	with
	\beq
		\hat{\Pi}_{hh}(p^2)&=& \frac{1}{(4\pi)^2}\int_{0}^{1} dx
	\Bigg\{
	e^2\Bigg[p^2(1-\ln\frac{m^2}{\m^2}-2\ln \frac{p^2 x(1-x)+m^2}{\m^2}) \nonumber\\
	&+&\frac{(m_h^4+2p^2m_h^2)}{2m^2} \ln \frac{p^2 x(1-x)+m_h^2}{\m^2}-6m^2(1-\ln \frac{m^2}{\m^2}+\ln \frac{p^2 x(1-x)+m^2}{\m^2})\Bigg]\nonumber\\
	&+&\lambda \Big[\frac{1}{2}m_h^2(-6+6\ln \frac{m_h^2}{\m^2}-9\ln \frac{p^2 x(1-x)+m_h^2}{\m^2})\Big]\nonumber\\
	&-&\Bigg[\xi (e^2p^2+\l m^2)(1- \ln \frac{\xi m^2}{\m^2})+(e^ 2\frac{(m_h^4+2p^2m_h^2)}{2m^2}+\lambda\frac{m_h^2}{2})\ln \frac{p^2 x(1-x)+\xi m^2}{\m^2}\Bigg]
	\Bigg\}.
	\label{dk6}
	\eeq
	and
	\beq
	C_{hh}(p^2)&=& - \frac{e^2}{2m^2(4\pi)^2}\int_{0}^{1} dx \Big\{ \ln\left( \frac{p^2 x(1-x)+m_h^2}{\m^2}\right)- \ln\left( \frac{p^2 x(1-x)+\xi m^2}{\m^2}\right)\Big\}
	\eeq
	and the   reliable resummed approximation becomes
	\beq
	G_{hh}(p^2)&=& \frac{1}{p^2+m^2-\hat{\Pi}(p^2)}+C_{hh}(p^2),
	\eeq
	which is shown in FIG.~\ref{prophhh}.
	
	For completeness, let us mention here that the integrals over the Feynman parameter $x$ that appear in the propagators can be done analytically, see Appendix \ref{app}. Since the transverse component $A^T_\mu$ of the Abelian gauge field is gauge-invariant, it turns out that the transverse photon propagator is independent from the gauge parameter $\xi$,  while the Higgs propagator does depend on $\xi$, in agreement with the Nielsen identities  analyzed in \cite{Haussling:1996rq}.
	
	\subsection{The correlation functions  $\langle O(x) O(y) \rangle$ and $ \langle V_{\mu}(x) V_{\nu}(y) \rangle$  at one-loop order \label{fmsp}}
	We are now ready to study the two-point correlation functions of the local gauge-invariant operators $(V_{\mu}(x), O(x))$. For the correlator of the scalar composite operator we get:	
	\beq
	\braket{{O}(x) {O}(y)} &=& v ^2 \braket{h(x) h(y)}+ v  \braket{h(x) \rho (y)^2}+ v  \braket{h(x)  h(y)^2}+ \nonumber \\
	&&+\frac{1}{4} \Big( \braket{h(x)^2 \rho (y)^2}+\braket{h(x)^2 h(y)^2}+ \braket{\rho (x)^2 \rho (y)^2}\Big). \label{exp1}
	\eeq
	Individually, the terms in the expansion \eqref{exp1} are not gauge-invariant, but their sum is. We can now analyze the connected diagrams for each term, up to one-loop order, through the action \eqref{fullaction2}. We calculated the one-loop diagrams in Appendix \ref{A}. Looking at the diagrams in FIG.~\ref{Yw}, we can see that the correlation function $\braket{O(p)O(-p)}$ will have the following structure
\beq
\braket{O(p)O(-p)}^{1-{\rm loop}} &=&  \frac{A_{\rm fin}(p^2)+\delta A_{\rm div}(p^2) }{(p^2+m_h^2)^2} + \frac{B_{\rm fin}(p^2)+\delta B_{\rm div}(p^2)}{(p^2+m_h^2)}\nonumber\\
&+& C_{\rm fin}(p^2)+\delta C_{\rm div}(p^2),
\label{finn}
\eeq
where $(A_{\rm fin}, B_{\rm fin}, C_{\rm fin})$ stand for the finite parts and $(\delta A_{\rm div}, \delta B_{\rm div}, \delta C_{\rm div})$ for the purely divergent terms, i.e.~the   one-loop pole terms in $\frac{1}{\epsilon}$ obtained by means of the  dimensional regularization ($d=4-\epsilon$), namely
\beq
\delta A_{\rm div}(p^2)&\overset{\epsilon \rightarrow 0}{=}&\frac{ v^2}{8\pi^2 \epsilon} \Big(2v^2 \lambda^2-e^2(p^2(-3+\xi)+v^2\lambda \xi)\Big),\nonumber \\
\delta B_{\rm div}(p^2)&\overset{\epsilon \rightarrow 0}{=}&\frac{v^2 (6 e^4-\lambda^2+e^2 \lambda \xi)}{8 \pi^2 \epsilon \lambda},\nonumber \\
\delta C_{\rm div}(p^2) &\overset{\epsilon \rightarrow 0}{=}& \frac{1}{8 \pi^2 \epsilon},
\eeq
while
\beq
A_{\rm fin} (p^2)&=& \frac{v^2}{(4\pi)^2}\int_{0}^{1} dx
	\Bigg\{
	e^2\Bigg[p^2(1-\ln\frac{m^2}{\m^2}-2\ln \frac{p^2 x(1-x)+m^2}{\m^2}) \nonumber\\
	&-&\frac{p^4}{2m^2} \ln \frac{p^2 x(1-x)+m^2}{\m^2}-6m^2(1-\ln \frac{m^2}{\m^2}+\ln \frac{p^2 x(1-x)+m^2}{\m^2})\Bigg]\nonumber\\
	&+&\lambda \Big[\frac{1}{2}m_h^2(-6+6\ln \frac{m_h^2}{\m^2}-9\ln \frac{p^2 x(1-x)+m_h^2}{\m^2})\Big]\nonumber\\
	&-&\Bigg[\xi (e^2p^2+\l m^2)(1- \ln \frac{\xi m^2}{\m^2})-(e^ 2\frac{p^4}{2m^2}-\lambda\frac{m_h^2}{2})\ln \frac{p^2 x(1-x)+\xi m^2}{\m^2}\Bigg]\Bigg\},\nonumber \\
B_{\rm fin}(p^2)&=&  \frac{1}{(4\pi)^2m_h^2}\int_{0}^{1} dx \Bigg\{-m^2 \xi  m_h^2 \ln \left(\frac{m^2 \xi }{\mu ^2}\right)+m^2 \xi  m_h^2+m_h^4 \Big(3 \ln \left(\frac{ m_h^2+p^2 (1-x) x}{\mu ^2}\right)\nonumber \\
&+&\ln \left(\frac{m^2 \xi+p^2 (1-x) x}{\mu ^2}\right)\Big)-3 m_h^4 \ln \left(\frac{m_h^2}{\mu ^2}\right)+3 m_h^4+2 m^4-6 m^4 \ln \left(\frac{m^2}{\mu ^2}\right)\Bigg\},\nonumber \\
C_{\rm fin}(p^2) &=& -\frac{1}{2(4\pi)^2}\int_{0}^{1} dx \Bigg\{\ln \left(\frac{m_h^2+p^2 (1-x) x}{\mu ^2}\right)+\ln \left(\frac{m^2 \xi +p^2 (1-x) x}{\mu ^2}\right)\Bigg\}.
\eeq
The divergent terms $(\delta A_{\rm div}, \delta B_{\rm div}, \delta C_{\rm div})$ can be eliminated by means of the   standard counterterms as well as by suitable counterterms in the external   source part of the action $S_J$ accounting for the introduction of the composite operator $O(x)$, see \cite{KlubergStern:1974rs, Piguet:1995er,itzykson2012quantum, Joglekar:1975nu,Dudal:2008tg} for a general account on this topic, i.e.
\beq
S_J&=& S +
\int d^4 x \Bigg[ ( 1+\delta Z_{\rm div}^0) J(x) O(x) + ( 1+\delta Z_{\rm div}) \frac{(J(x))^2}{2}\Bigg],
\label{1o}
\eeq
where $J(x)$ is a BRST invariant dimension two source needed to define the generator $Z^c(J)$ of the connected Green function $\braket{O(x)O(y)}$:
\beq
\braket{O(x)O(y)}=\left.\frac{\delta^2 Z^c(J)}{\delta J(x) \delta J(y)}\right|_{J=0}.
\eeq
It is worth emphasizing here that we have the freedom of introducing a pure   BRST invariant contact  term in the external source $J(x)$:
\beq
\int d^4 x \,\frac{\alpha}{2}\,J^2(x),
\label{3d}
\eeq
which can be arbitrarily added to the action \eqref{1o}. Including such a term in \eqref{1o} will have the effect of adding a dimensionless constant to $G_{OO}=\braket{O(p)O(-p)}$, i.e.
\beq
G_{OO}(p^2)\rightarrow G_{OO}(p^2)+\alpha
\label{3e}.
\eeq
In particular, $\alpha$ can be chosen to be equal to $-G_{OO}(0)$, implying then that the modified Green's function
\beq
G_{OO}(p^2)-G_{OO}(0)
\label{dv1}
\eeq
will obey a   once substracted KL representaion, see Section \ref{IIII}   for more details on this.

Inserting the unity
\beq
1= (p^2+m_h^2)/(p^2+m_h^2) = ((p^2+m_h^2)/(p^2+m_h^2))^2,
\label{2o}
\eeq
into the finite part of $\braket{O(p)O(-p)}$, we write
	\beq
	\langle {O}(p) {O}(-p) \rangle_{\rm fin} &=& \frac{v^2}{p^2+m_h^2}+\frac{\hbar v^2}{(p^2+m_h^2)^2}\Pi (p^2)+\mathcal{O}(\hbar^2)
	\eeq
	where
	\beq
	\Pi_{OO}(p^2) &=&  \frac{1}{v^2}\Big((A_{\rm fin}(p^2)) + (p^2+m_h^2)(B_{\rm fin}(p^2))+ (C_{\rm fin}(p^2)) (p^2+m_h^2)^2\Big), \nonumber
\\
	&=&\frac{1}{32 \pi ^2  v^2 m_h^2}
	\int_0^1 dx \Bigg\{-8 m_h^2 m^4-2 m^2 p^2 (m_h^2+6 m^2) \ln \left(\frac{m^2}{\mu ^2}\right)+
	\nonumber\\
	&& + m_h^2 \Big[-(p^2-2 m_h^2)^2 \ln \left(\frac{m_h^2+p^2(1-x) x}{\mu ^2}\right)
	%\nonumber \\
	%&&
	-(12 m^4+4 m^2 p^2+p^4) \ln \left(\frac{m^2+p^2(1-x) x}{\mu ^2}\right)\Big]+
	\nonumber \\
		&&+2 p^2 (3 m_h^4+m_h^2 m^2+2 m^4)-6 m_h^4 p^2 \ln \left(\frac{m_h^2}{\mu ^2}\right)\Bigg\}.
			\label{ark}
	\label{dk3}
	\eeq
Since \eqref{ark} contains terms of the order of $\frac{p^4}{p^2+m^2}\ln (p^2)$,
we follow the steps \eqref{jju1}-\eqref{jju2} to find the resummed   propagator in the one-loop approximation
	\beq
	G_{{O}{O}}(p^2)
	&=&\frac{v^2}{p^2+m_h^2-\hat{\Pi}_{OO}(p^2)}+C_{OO}(p^2)
	\label{dk3}
	\eeq
	with
	\beq
	\hat{\Pi}_{OO}(p^2)&=&\frac{1}{32 \pi ^2  v^2 m_h^2}
	\int_0^1 dx \Bigg\{-8 m_h^2 m^4-2 m^2 p^2 (m_h^2+6 m^2) \ln \left(\frac{m^2}{\mu ^2}\right)+
	\nonumber\\
	&& \hspace{-0.7cm} +~ m_h^2 \Big[3(m_h^4+2m_h^2p^2) \ln \left(\frac{m_h^2+p^2(1-x) x}{\mu ^2}\right)
	%\nonumber \\
	%&&
	-(12 m^4+4 m^2 p^2-m_h^4-2p^2m_h^2) \ln \left(\frac{m^2+p^2(1-x) x}{\mu ^2}\right)\Big]+
	\nonumber \\
		&&~+2 p^2 (3 m_h^4+m_h^2 m^2+2 m^4)-6 m_h^4 p^2 \ln \left(\frac{m_h^2}{\mu ^2}\right)\Bigg\},
	\label{dk3}
	\eeq
	and
	\beq
	C_{OO}(p^2)&=&-\frac{1}{32 \pi ^2  }
	\int_0^1 dx \Bigg\{\ln \left(\frac{m_h^2+p^2x(1-x)}{\mu^2}\right)+\ln \left( \frac{ m^2 + p^2 x (1-x) }{\mu^2}\right)\Bigg\}.
	\label{grgr}
	\eeq
  It is crucial to stress here that, if we had also resummed $C_{OO}(p^2)$ into the inverse propagator, we would have encountered an (unphysical) tachyon into the composite operator propagator, as at some point the exploding large $p^2$-behaviour in $C_{OO}(p^2)$ would completely wash out the other ``UV tamed'' contributions.

\begin{figure}[t]
			\center
			\includegraphics[width=12cm]{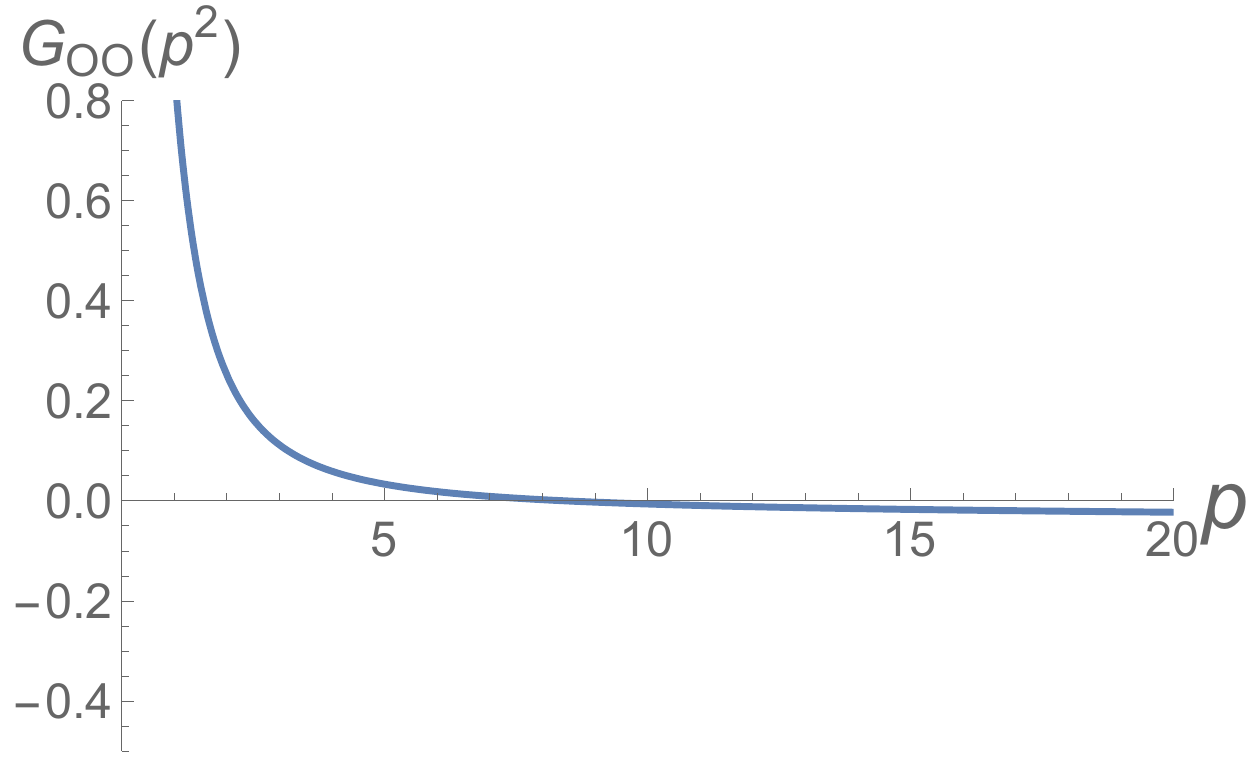}
			\caption{Resummed   propagator for the scalar composite operator.
			All quantities are given in units of appropriate powers of the energy scale $\mu$, with the parameter values $e=1$, $v=1 \, \mu$, $\lambda=\frac{1}{5}$.}
			\label{prophh1}
		\end{figure}

	The   propagator is depicted in FIG.~\ref{prophh1} and we notice that the Green function $G_{OO}(p^2)$ becomes negative for large enough values of the momentum $p$. As one realizes from expression \eqref{dk3}, this feature is due to the growing in the UV region of the logarithms contained in the term $C_{OO}(p^2)$, see eq.~\eqref{grgr}. It is worth mentioning that this behaviour is also present when the parameter $v$ is completely removed from the theory. In fact, setting $v=0$, the action $S_0$ in eq.~\eqref{higgsqed}, reduces to that of massless scalar QED, namely
\beq
\label{dd2}
  \left.S_0\right|_{v=0}=\int d^4x \left(\frac{F_{\mu\nu}^2}{4}+ (D_\mu\varphi)^\dagger(D_\mu\varphi) + \frac{\lambda}{2}(\varphi^\dagger\varphi)^2\right)\,,
\eeq
with
\beq
\label{dd3}
  \left.\varphi\right|_{v=0}=\frac{1}{\sqrt 2}(h+i\rho).
\eeq
Of course, when $v=0$, the operators $O=\varphi^\dagger \varphi$ and $V_\mu=-i \varphi^\dagger D_\mu \varphi$ are still gauge-invariant. Though, from eqs.~\eqref{dk3}-\eqref{grgr}, computing $\braket{O(p)O(-p)}_{v=0}$, one immediately gets
\begin{equation}\label{dd1}
  \braket{O(p)O(-p)}_{v=0}=\left.C_{OO}(p^2)\right|_{v=0}=-\frac{1}{16\pi^2} \int_0^1 dx \ln \frac{p^2x(1-x)}{\mu^2}.
\end{equation}
This equation precisely shows that the term $C_{OO}(p^2)$, and thus the negative behaviour for large enough values of $p$, is what one usually obtains in a theory for which $v=0$, making evident that the presence of $C_{OO}(p^2)$ is not peculiarity of the $U(1)$ Higgs model, on the contrary. However, in addition to the term $C_{OO}(p^2)$ and unlike massless scalar QED, the correlation function $\braket{O(p)O(-p)}$ of the $U(1)$ Higgs model exhibits the term $\frac{v^2}{p^2+m_h^2-\hat\Pi_{OO}}$, which will play a pivotal role. Indeed, as we shall see later on, this term, originating from the expansion of $\varphi$ around the minimum of the Higgs potential, $\varphi=\frac{1}{\sqrt 2}(v+h+i\rho)$, will enable us to devise a gauge-invariant description of the elementary excitations of the model.

\begin{figure}[t]
			\center
			\includegraphics[width=12cm]{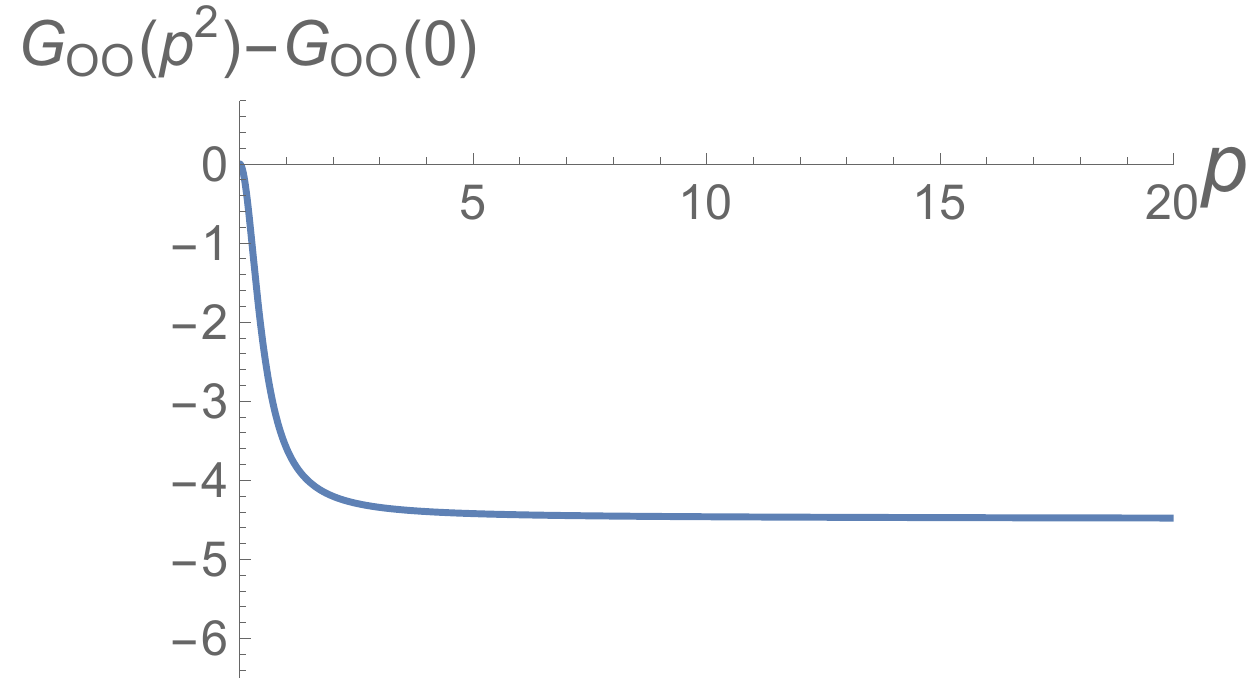}
			\caption{The resummed   propagator with a single subtraction.
		All quantities are given in units of appropriate powers of the energy scale $\mu$, with the parameter values $e=1$, $v=1 \, \mu$, $\lambda=\frac{1}{5}$.}
			\label{prophh2}
		\end{figure}

Let us end the analysis of the correlation function $G_{OO}(p^2)$ by displaying the behaviour of its first derivative, $\frac{\partial G_{OO}(p^2)}{\partial p^2}$, as well as of the   once subtracted correlator $G_{OO}(p^2) - G_{OO}(0)$,
 see FIG.~\ref{prophh2}. The first derivative, as expected, is negative while, unlike $G_{OO}(p^2)$,   it decays to zero for $p^2\to\infty$. The quantity $\frac{\partial G_{OO}(p^2)}{\partial p^2}$ will be helpful when discussing the spectral representation corresponding to $\braket{O(p)O(-p)}$.
			\begin{figure}[t]
			\center
			\includegraphics[width=12cm]{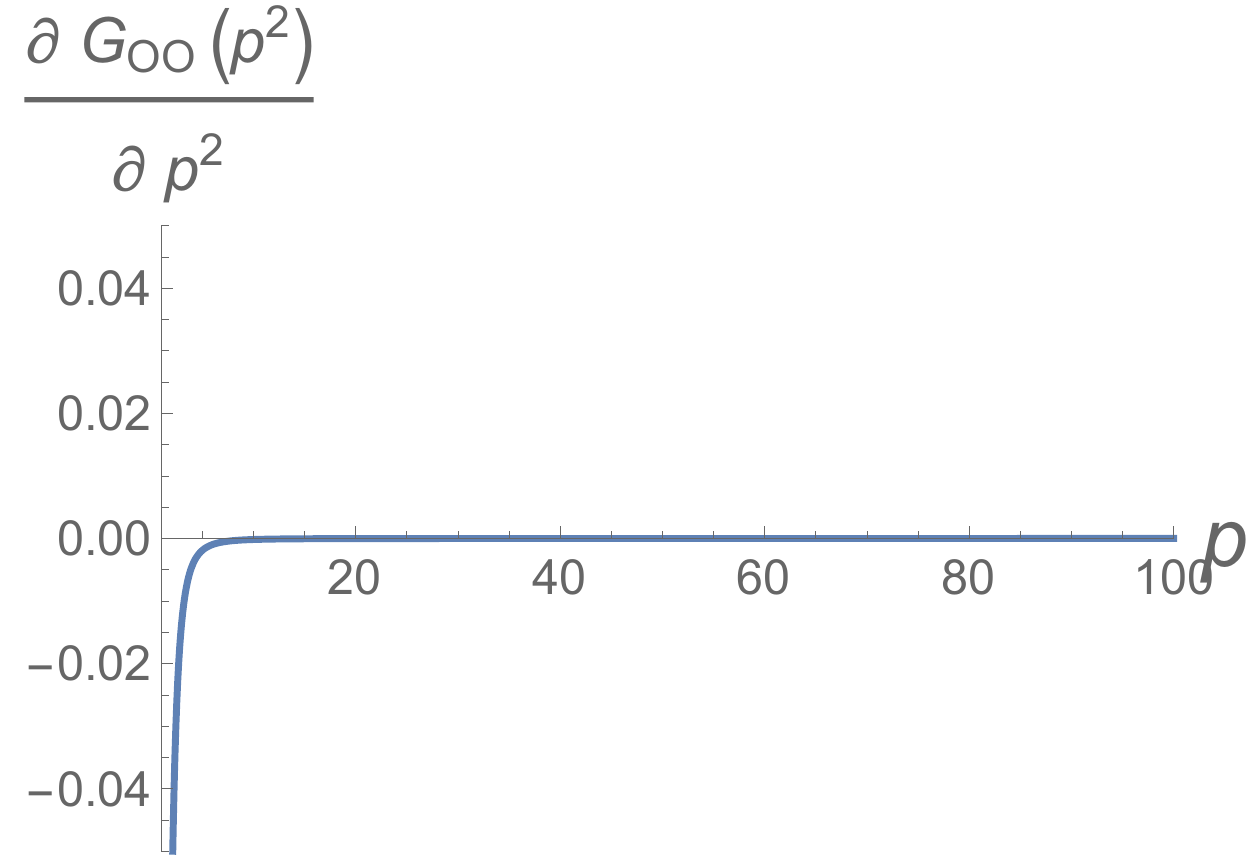}
			\caption{The first derivative of the   resummed propagator.
		All quantities are given in units of appropriate powers of the energy scale $\mu$, with the parameter values $e=1$, $v=1 \, \mu$, $\lambda=\frac{1}{5}$. }
			\label{prophh3}
		\end{figure}

\begin{figure}[t]
			\center
			\includegraphics[width=12cm]{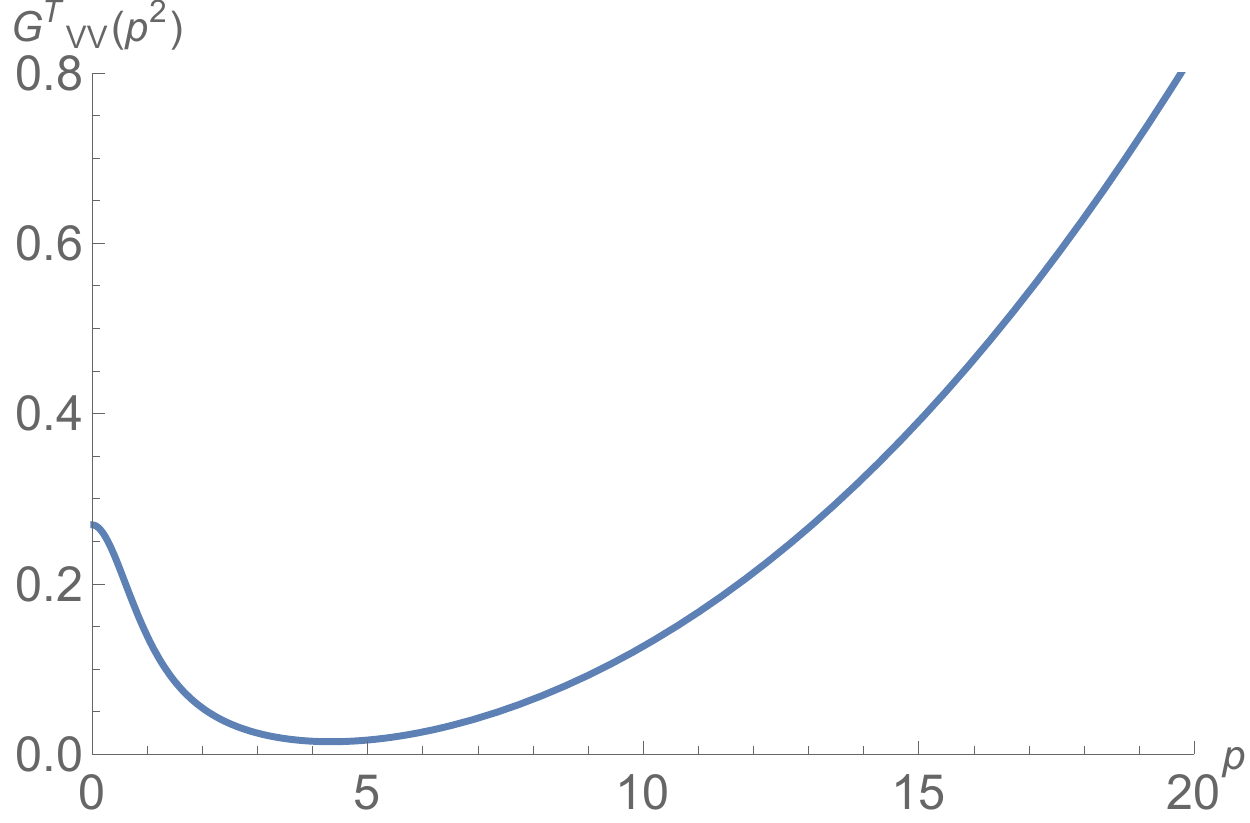}
			\caption{  Resummed propagator for the vector composite operator. All quantities are given in units of appropriate powers of the energy scale $\mu$, with the parameter values $e=1$, $v=1 \, \mu$, $\lambda=\frac{1}{5}$.}
			\label{propAAT1}
		\end{figure}

\begin{figure}[t]
			\center
			\includegraphics[width=12cm]{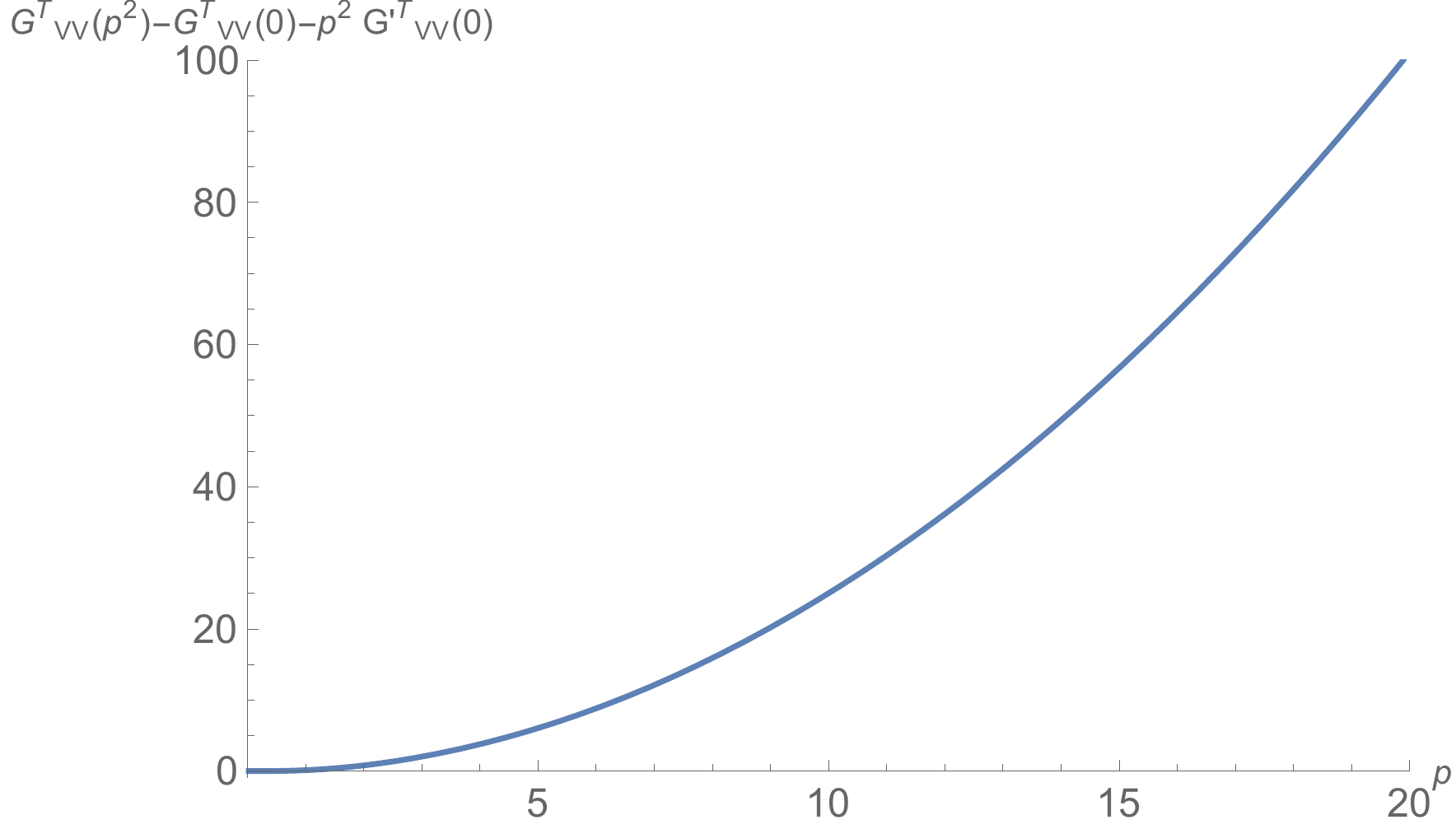}
			\caption{  Resummed propagator with a double subtraction for the vector composite operator.
		All quantities are given in units of appropriate powers of the energy scale $\mu$, with the parameter values $e=1$, $v=1 \, \mu$, $\lambda=\frac{1}{5}$.}
			\label{propAAT1b}
		\end{figure}

\begin{figure}[t]
			\center
			\includegraphics[width=12cm]{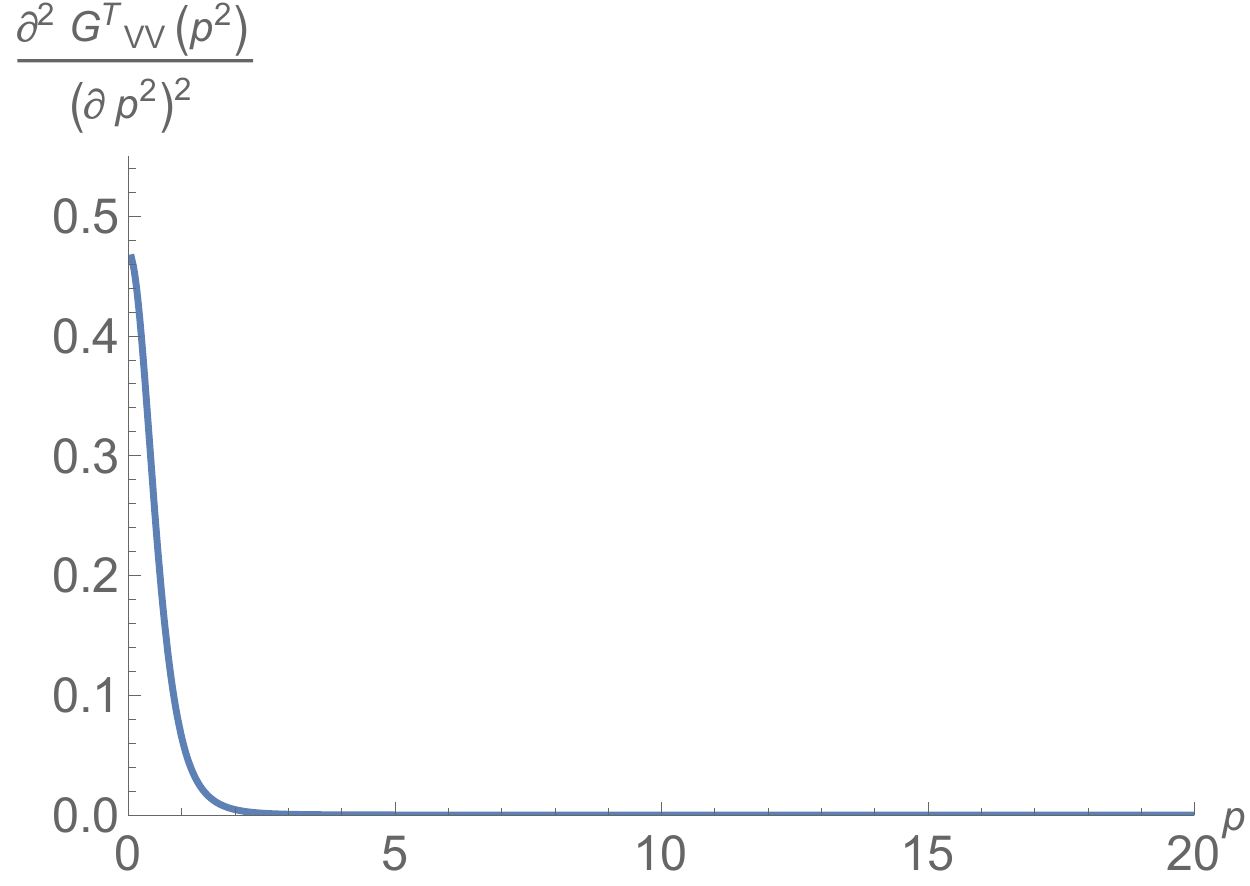}
			\caption{  Second derivative of the vector propagator.
		All quantities given in units of appropriate powers of the energy scale $\mu$, with the parameter values $e=1$, $v=1 \, \mu$, $\lambda=\frac{1}{5}$.}
			\label{propAAT1c}
		\end{figure}

Then, for the vectorial composite operator ${V}_{\mu}(x)$, we first observe that
	\beq
	V_{\mu}(x)&=& -i \varphi^{\dagger}(x) (D_{\mu}\varphi)(x)\nonumber \\
	&=& e \vf^{\dagger}(x) A_{\mu}(x) \vf(x) -\frac{1}{2} i\varphi^{\dagger}(x)\partial_{\mu} \vf(x) +\frac{1}{2}i \varphi(x)\partial_{\mu} \vf^{\dagger}(x)- i \partial_{\mu} O(x),
	\eeq
	and since we know that the last term is gauge-invariant, the first three terms together   must also be.  We can thus define a new gauge-invariant operator
	\beq
	V'_{\mu}(x)
	&=& e \vf^{\dagger}(x) A_{\mu}(x) \vf(x) -\frac{1}{2} i\varphi^{\dagger}(x)\partial_{\mu} \vf(x) +\frac{1}{2}i \varphi(x)\partial_{\mu} \vf^{\dagger}(x),
	\eeq
	expanding the scalar field $\vf(x)$ we find
		\beq
		V'_{\mu}(x)
	&=& \frac{1}{2} \Big( e (v+h(x))^2 A_{\mu}(x) + e \rho^2(x) A_{\mu}(x)+(v+h(x))\partial_{\mu} \rho (x)-\rho(x) \partial_{\mu} h(x) \Big)
	\eeq
	so that
	\beq
	  \braket{V'_{\m}(x)V'_{\n}(y)} &\overset{\varphi \rightarrow \frac{1}{\sqrt{2}}(v + h +i \rho)}{=}&-\frac{1}{4}\Bigg\{-e^2 v^4  \langle A_{\mu }(x) A_{\nu}(y) \rangle-4 e^2 v^3  \langle h(x)  A_{\mu }(x)A_{\nu}(y) \rangle \nonumber \\
	&&-2 e^2 v^2  \langle h(x)^2A_{\mu }(x)  A_{\nu}(y) \rangle-4 e^2 v^2  \langle h(x)  A_{\mu }(x) h(y) A_{\nu}(y)\rangle \nonumber \\
	&&-2 e^2 v^2 \langle \rho (x)^2A_{\mu }(x) A_{\nu}(y)  \rangle-2 e v^2 \partial ^x_{\mu } \langle h(x) \rho (x) A_v(y) \rangle \nonumber \\
	&&+4 e v^2  \langle \partial ^x_{\mu } h(x)\rho (x)A_{\nu}(y) \rangle -2 e v^3\partial ^x_{\mu } \langle \rho (x) A_{\nu}(y) \rangle\nonumber \\
	&&-4 e v^2 \partial ^x_{\mu } \langle  \rho (x) h(y) A_v(y) \rangle-2 v \partial ^x_{\mu }\partial ^y_{\nu } \langle h(x) \rho (x) \rho (y) \rangle \nonumber \\
	&&+4 v \partial ^y_{\nu } \langle \partial ^x_{\mu }h(x) \rho (x)\rho (y) \rangle-\partial ^x_{\mu }\partial ^y_{\nu } \langle h(x) \rho (x) h(y) \rho (y) \rangle \nonumber \\
	&&+4 \langle\partial ^x_{\mu }h(x)\rho (x) h(y) \partial ^y_{\nu }\rho (y) \rangle-v^2\partial ^x_{\mu }\partial ^y_{\nu }  \langle\rho (x) \rho (y) \nonumber \rangle \Bigg\}\nonumber \\
	&&+O(\hbar^2),
	\label{exp2}
	\eeq
	where we have discarded the terms that do not have one-loop contributions.
	In momentum space, we can split the two-point function into transverse and longitudinal parts in the usual way:
	\beq
	  \langle V'_{\mu}(p)V'_{\nu}(-p)\rangle =\langle {V'}(p) {V'}(-p)\rangle^T \mathcal{P}_{\mu\nu}+\langle V'(p)V'(-p)\rangle^L \mathcal{L}_{\mu\nu} \,,
	\eeq
	where we have introduced the transverse and longitudinal projectors, given respectively by
	\beq
	  \mathcal{P}_{\mu\nu}(p)&=& \delta_{\mu\nu}-\frac{p_{\mu}p_{\nu}}{p^2}\, \qquad	\mathcal{L}_{\mu\nu}(p)~=~ \frac{p_{\mu}p_{\nu}}{p^2}\,.
	\eeq
		At tree-level, we find in momentum space
	\beq
	\langle V'_{\mu}(p)V'_{\nu}(-p)\rangle_{\text{\text{tree}}}&=&-\frac{1}{4}\left(-e^2 v^4  \langle  A_{\mu }(p)A_{\nu}(-p) \rangle-v^2p_{\mu }p_{\nu }  \langle\rho (p) \rho (-p) \rangle\right) \nonumber \\
	&=&\frac{1}{4}\left(e^2 v^4\frac{1}{p^2+m^2} \mathcal{P}_{\mu\nu}+e^2 v^4\frac{\xi}{p^2+\xi m^2}\mathcal{L}_{\mu \nu}+ v^2 \frac{p^2}{p^2+\xi m^2}\mathcal{L}_{\mu \nu}\right) \nonumber \\
	&=&\frac{e^2 v^4}{4}\frac{1}{p^2+m^2} \mathcal{P}_{\mu\nu}+v^2\mathcal{L}_{\mu \nu}.
	\eeq
		We can now analyze the connected diagrams for each term, up to one-loop order, through the action \eqref{fullaction2}. We calculated the one-loop diagrams in Appendix \ref{ah}. Let us start with the transverse part. Looking at the diagrams in FIG.~\ref{Y22}, we can see that the one-loop correlation function will have the following structure
\beq
\braket{V'(p)V'(-p)}^{T,1-{\rm loop}} &=&  \frac{A^V_{\rm fin}(p^2)+\delta A^V_{\rm div}(p^2)}{(p^2+m^2)^2} + \frac{B^V_{\rm fin}(p^2)+\delta B^V_{\rm div}(p^2)}{(p^2+m^2)}\nonumber \\
&+& C^V_{\rm fin}(p^2)+\delta C^V_{\rm div}(p^2)
\label{finn}
\eeq
where $(A^V_{\rm fin}, B^V_{\rm fin}, C^V_{\rm fin})$ stand for the finite parts and $(\delta A^V_{\rm div}, \delta B^V_{\rm div}, \delta C^V_{\rm div})$ for the purely divergent terms, i.e.~the   one-loop pole terms in $\frac{1}{\epsilon}$ obtained by means of the  dimensional regularization, namely
\beq
\delta A^V_{\rm div}&\overset{\epsilon \rightarrow 0}{=}&\frac{e^4 v^4}{2 (4\pi)^2 \epsilon} \Big(\frac{1}{3}p^2-6(\frac{e^2}{\lambda}-\frac{1}{2})e^2 v^2+3\lambda v^2\Big),\nonumber \\
\delta B^V_{\rm div}&\overset{\epsilon \rightarrow 0}{=}&\frac{v^2 }{(4 \pi)^2 \epsilon}(6 \frac{e^6v^2}{\lambda}-3 e^4 v^2-\frac{e^2p^2}{3}+3e^2 \lambda v^2),\nonumber \\
\delta C^V_{\rm div} &\overset{\epsilon \rightarrow 0}{=}& \frac{1}{6 (4\pi)^2 \epsilon}(9e^2v^2-p^2-3\lambda v^2)
\eeq
and
\begin{allowdisplaybreaks}
\beq
A^V_{\rm fin} &=&\frac{e^4v^4}{2(4\pi)^2}\int_{0}^{1} dx \,\,\Bigg\{p^2 x(1-x)+m^2x \nonumber\\
	&+&m_h^2(1-x)(1-\ln\frac{p^2 x(1-x)+m^2x+m_h^2(1-x)}{\m^2})+m_h^2(1-\ln\frac{m_h^2}{\m^2})\nonumber\\
	&+&\frac{m^4}{m_h^2}(1-3 \ln \frac{m^2}{\m^2})+2m^2 \ln \frac{p^2 x(1-x)+m^2x+m_h^2(1-x)}{\m^2}\Bigg\},\nonumber \\
B^V_{\rm fin}&=& \frac{m^2}{18 m_h^2 p^2 (4\pi)^2} \int_0^1 dx   \Bigg\{3 m_h^4 \left(m_h^2-m^2-7 p^2\right) \ln \left(\frac{m_h^2}{\mu ^2}\right),\nonumber\\
&-&3 m_h^2 \left(2 p^2 \left(m_h^2-5 m^2\right)+\left(m_h^2-m^2\right)^2+p^4\right) \ln \left(\frac{x m_h^2+m^2 (1-x)+p^2 (1-x) x}{\mu ^2}\right)\nonumber\\
&-&3 \left(m_h^3-m^2 m_h\right)^2+9 p^2 \left(m^2 m_h^2+3 m_h^4+2 m^4\right)+2 p^4 m_h^2\nonumber\\
&-&3 m^2 \left(m_h^2 \left(p^2-m^2\right)+m_h^4+18 m^2 p^2\right) \ln \left(\frac{m^2}{\mu ^2}\right)\Bigg\}\nonumber \\
C^V_{\rm fin} &=& \frac{1}{36 (4 \pi )^2 p^2}\int_0^1 dx \Bigg\{3 m^2 \left(m_h^2-m^2+p^2\right) \ln \left(\frac{m^2}{\mu ^2}\right)\nonumber \\
&+&3 m_h^2 \left(-m_h^2+m^2+p^2\right) \ln \left(\frac{m_h^2}{\mu ^2}\right)+6 m_h^2 \left(p^2-m^2\right)-5 p^2 \left(3 m_h^2-9 m^2+p^2\right)\nonumber \\
&+&3 \left(2 m_h^2 \left(p^2-m^2\right)+m_h^4+m^4-10 m^2 p^2+p^4\right) \ln \left(\frac{ p^2 x(1-x)+xm^2+(1-x) m_h^2}{\mu ^2}\right)\nonumber \\
&+& 3 m_h^4+3 m^4-54 m^2 p^2+3 p^4\Bigg\}.
\eeq
\end{allowdisplaybreaks}
 The divergent terms $(\delta A^V_{\rm div}, \delta B^V_{\rm div}, \delta C^V_{\rm div})$ can   again be eliminated by means of the standard counterterms as well as by suitable counterterms in the external source part of the action $S^V_J$ accounting for the introduction of the composite operator $V'_{\m}(x)$, i.e.
\beq
S^V_J&=& S +
\int d^4 x \Bigg[ ( 1+\delta Z_{\rm div}^{V,0}) J_{\m}(x) V_{\m}(x) + ( 1+\delta Z^V_{\rm div}) \frac{J_{\m}(x)J_{\m}(x)}{2}\Bigg],
\label{1o1}
\eeq
where $J_{\m}(x)$ is a BRST invariant dimension one source needed to define the generator $Z^c(J)$ of the connected Green function $\braket{V'_{\m}(x)V'_{\n}(y)}$:
\beq
\braket{V'_{\m}(x)V'_{\n}(y)}=\left.\frac{\delta^2 Z^c(J)}{\delta J_{\m}(x) \delta J_{\n}(y)}\right|_{J=0},
\eeq
and like in the scalar case, we have the freedom of introducing   BRST invariant pure contact terms in the external source $J_{\m}(x)$:
\beq
\int d^4 x \,\frac{1}{2}(\b v^2 \,J_{\m}(x)J_{\m}(x)+\gamma J_{\m}(x)\partial^2 J_{\m}(x)+\sigma (\partial_{\m}J_{\m}(x))^2),
\label{3dd}
\eeq
which can be arbitrarily added to the action eq.~\eqref{1o1}. Including   such terms in \eqref{1o1} will have the effect of adding   a first order polynomial in $p^2$ to $G^T_{VV}(p^2)=\braket{V'(p)V'(-p)}^T$, i.e.
\beq
G^T_{VV}(p^2)\rightarrow G^T_{VV}(p^2)+\beta v^2+\gamma p^2
\label{3e},
\eeq
where we notice that the last term in \eqref{3dd} does not contribute to the transversal part of the propagator.
In particular, $\beta$ and $\gamma$ can be   chosen so that \eqref{3e} becomes
\beq
G^T_{VV}(p^2)-G^T_{VV}(0)-p^2\left. \frac{\partial G^T_{VV}(p^2)}{\partial p^2}\right|_{p=0}.
\label{dv2}
\eeq
  Eventually, we have a Green's function that obeys a twice substracted KL representation, see Section \ref{IIII}.
Following   similar steps as for the scalar composite field, \eqref{1o}-\eqref{2o}, we find
	\beq
	\langle V'(p)V'(-p)\rangle^T&=&\frac{e^2 v^4}{4}\frac{1}{p^2+m^2}+ \frac{\hbar e^2 v^4}{4}\frac{\Pi^T_{VV}(p^2)}{(p^2+m^2)^2}+ \mathcal{O}(\hbar^2),
\eeq
with
	\beq
	\Pi^T_{VV}(p^2)
	&=&-\frac{1}{9 (4\pi)^2 e^2 v^4 m_h^2}\int_0^1 dx\Bigg\{-18 m^4 (m_h^4+m^4)+9 m_h^2 p^4 (m_h^2+m^2) \nonumber \\
		&-&3 m_h^2 p^2 \big[2 p^2 (m_h^2-5 m^2)+(m_h^2-m^2)^2+p^4\big] \ln \left(\frac{m_h^2 (1-x)+m^2 x+p^2 (1-x) x}{\mu ^2}\right)+2 m_h^2 p^6 \nonumber \\
		&+&3 m_h^4 \ln \left(\frac{m_h^2}{\mu ^2}\right) \big[p^2 (m_h^2+11 m^2)+6 m^4-p^4\big] \nonumber \\
		&+&3 m^2 \big[-m_h^2 p^4+p^2 (-m_h^4+m_h^2 m^2+36 m^4)+18 m^6\big] \ln \left(\frac{m^2}{\mu ^2}\right) \nonumber \\
		&-&3 p^2 (m_h^6+10 m_h^4 m^2+m_h^2 m^4+12 m^6)\Bigg\},
		\label{dk4}
		\eeq
		and following the steps \eqref{jju1}-\eqref{jju2}, we find
			\beq
		G_{VV}^T&=& \frac{e^2 v^4}{4}\Big(\frac{1}{p^2+m^2-\hat\Pi^T_{VV}(p^2)}\Big)+ C_{VV}(p^2)
		\label{GAAA}
		\eeq
			with
			\beq
\hat{\Pi}^{T}_{VV}(p^2)
	&=&-\frac{1}{9 (4\pi)^2 e^2 v^4 m_h^2}\int_0^1 dx\Bigg\{-18 m^4 (m_h^4+m^4)+9 m_h^2 p^4 (m_h^2+m^2) \nonumber \\
		&-&3 m_h^2 \big[2 (-2m^2p^2-m^4) (m_h^2-5 m^2)+p^2(m_h^2-m^2)^2-2m^2p^2-m^4 \big] \ln \left(\frac{m_h^2 (1-x)+m^2 x+p^2 (1-x) x}{\mu ^2}\right) \nonumber \\
		&+&2 m_h^2 p^6+3 m_h^4 \ln \left(\frac{m_h^2}{\mu ^2}\right) \big[p^2 (m_h^2+11 m^2)+6 m^4-p^4\big] \nonumber \\
		&+&3 m^2 \big[-m_h^2 p^4+p^2 (-m_h^4+m_h^2 m^2+36 m^4)+18 m^6\big] \ln \left(\frac{m^2}{\mu ^2}\right) \nonumber \\
		&-&3 p^2 (m_h^6+10 m_h^4 m^2+m_h^2 m^4+12 m^6)\Bigg\},
		\label{dk4}
		\eeq
		and
		\beq
		C_{VV}(p^2)&=&\frac{1}{12 (4\pi)^2 }\int_0^1 dx\Bigg\{
		(2  m_h^2+p^2) \ln \left(\frac{m_h^2 (1-x)+m^2 x+p^2 (1-x) x}{\mu ^2}\right)\Bigg\}.
		\eeq
 	The resummed   propagator \eqref{GAAA} is depicted in FIG.~\ref{propAAT1}, as well as the subtracted version \eqref{dv2} in FIG.~\ref{propAAT1b} and the second derivative in FIG.~\ref{propAAT1c}, which will be   important for the spectral analysis in Section \ref{IIII}.
	
						
	
		For the longitudinal part of the propagator (see   Appendix~\ref{ah} for details), we find the divergent part
			\beq
		\langle V'(p)V'(-p)\rangle^L_{\rm div}&\overset{\epsilon \rightarrow 0}{=}& -\frac{v^2 \left(3 e^4+\lambda ^2\right)}{(4 \pi) ^2 \lambda  \epsilon }
		\eeq
		and the total finite correction up to first order in $\hbar$ is given by
		\beq
		\langle V'(p)V'(-p)\rangle^L_{\rm fin}&=& v^2	-\Bigg(\frac{m_h^4-m_h^4 \ln \left(\frac{m_h^2}{\mu ^2}\right)+m^4-3 m^4 \ln \left(\frac{m^2}{\mu ^2}\right)}{32 \pi ^2 m_h^2} \Bigg).
	\eeq
From this expression one sees that, as in the case of the tree level, the one-loop 
correction to the longitudinal part of the correlator $\braket{V'_{\mu} (x) V'_{\nu}(y)}^L_{\rm fin}$ 
remains independent from the momentum $p^2$. As such, it is not associated to any physical mode.

		\section{Spectral properties of the gauge-invariant local  operators $(V_{\mu}(x),O(x))$ \label{IIII}}
		In this section, we will   study the spectral properties associated with the correlation functions obtained in the last section. In \ref{2a}, we will shortly review the techniques employed in  \cite{us} to obtain the pole mass, residue and spectral density up to first order in $\hbar$. In \ref{2b}, we analyze the spectral properties of the elementary   propagators. In \ref{2c}, the spectral properties of the composite operators $(V_{\mu}(x),O(x))$ are discussed.

		\subsection{Obtaining the spectral function \label{2a}}
		For elementary fields we obtain the spectral density function by comparing the (  Euclidean) KL  spectral representation for the propagator of a generic field $\widetilde{O}(p)$
		\beq
		\braket{\widetilde{O}(p)\widetilde{O}(-p)}=G(p^2)=\int_0^{\infty} dt \frac{\rho (t)}{t+p^2},
		\label{ltt}
		\eeq
		where $\rho(t)$ is the spectral density function and $G(p^2)$ stands for the  resummed propagator
		\beq
		G(p^2)&=& \frac{1}{p^2+m^2-\Pi(p^2)}.
		\label{iuh}
		\eeq
		For higher-dimensional operators, the resummed propagator acquires an overall (dimensionful) factor identical to the one appearing in its tree level result, as we have seen in section \ref{fmsp}.
		 We also note that in the case of higher dimensional operators, the spectral representation,
			eq.\eqref{ltt}, might require appropriate subtraction terms in order to ensure   a convergent spectral integral. A standard way to cure this problem is to
			subtract from $G(p^2)$ the first few   (divergent) terms of its Taylor expansion at $p=0$ \cite{colangelo2001qcd}, making the integral more and more convergent. These subtraction terms are directly related to the renormalization of the composite operators, and one can see that the modified Green's functions for the composite scalar field \eqref{dv1} and for the composite vector field \eqref{dv2} are in fact subtractions of the Taylor series to first and second order, respectively. In our theory we can make use of the subtracted equations at $p=0$ because all fields are massive in the $R_{\xi}$-gauge, so there are no divergences at zero momentum. Also, we stress that the spectral function $\rho(t)$ is not affected by the subtraction procedure   as polynomials do not display discontinuities in the complex $p^2$-plane, whilst the spectral function is proportional to the jump across the branch cut. Moreover, we can see that these subtractions do not have an influence either on the (second) derivative of the propagator. For the scalar composite operator
			\beq
			\frac{\partial (G_{OO}(p^2)-G_{OO}(0))
			}{\partial p^2}= \frac{\partial G_{OO}(p^2)}{\partial p^2}=-\int_0^{\infty} dt \frac{\rho (t)}{(t+p^2)^2},
			\eeq
			which means that for a positive spectral function,   the first derivative of $G_{OO}(p^2)$ ought to be strictly negative, as is indeed confirmed from FIG.~\ref{prophh3}. For the vector composite operator
				\beq
			\frac{\partial^2 (G_{VV}(p^2)-G_{VV}(0)-p^2 G'_{VV}(0))
			}{(\partial p^2)^2}= \frac{\partial^2 G_{VV}(p^2)}{(\partial p^2)^2}= 2\int_0^{\infty} dt \frac{\rho (t)}{(t+p^2)^3},
			\eeq
		which should be strictly positive for a positive spectral function,   consistent with FIG.~\ref{propAAT1c}.

		We can also obtain the spectral function directly in the following way. The pole mass for any massless or massive field excitation is obtained by calculating the pole of the resummed propagator, that is, by solving
		\beq
		p^2+m^2-\Pi(p^2)=0 \, \label{ppp}
		\eeq
		and its solution defines the pole mass $p^2=-m_{\rm pole}^2$. As consistency requires us to work up to a fixed order in perturbation theory, we should solve eq. \eqref{ppp} for the pole mass in an iterative fashion. Therefore, to first order in  $\hbar$ , we find
		\beq
		m_{\rm pole}^2=m^2-\Pi^{1-{\rm loop}}(-m^2)+\mathcal{O}(\hbar^2),
		\label{pol}
		\eeq
		where $\Pi^{1-{\rm loop}}$ is the first order, or one-loop, correction to the propagator. Now, we write eq.\eqref{iuh} in a slightly different way, namely
		\beq
		G(p^2)&=& \frac{1}{p^2+m^2-\Pi(p^2)}\nonumber\\
		&=&\frac{1}{p^2+m^2-\Pi^{1-{\rm loop}}(-m^2)-(\Pi(p^2)-\Pi^{1- {\rm loop}}(-m^2))}\nonumber\\
		&=&\frac{1}{p^2+m_{\rm pole}^2-\widetilde{\Pi}(p^2)},
		\label{op}
		\eeq
		where we defined $\widetilde{\Pi}(p^2)=\Pi(p^2)-\Pi^{1-{\rm loop}}(-m^2)$. At one-loop, expanding  $\widetilde{\Pi}(p^2)$ around $p^2=-m^2_{\rm pole}=-m^2+\mathcal{O}(\hbar)$  gives the residue
		\beq
		Z&=& \lim_{p^2 \rightarrow -m_{\rm pole}^2} (p^2+m_{\rm pole}^2)G(p^2) \nonumber \\
		&=&\frac{1}{1-\partial_{p^2} \Pi(p^2)\vert_{p^2=-m^2}} \nonumber \\
		&=&1+\partial_{p^2} \Pi(p^2)\vert_{p^2=-m^2}+\mathcal{O}(\hbar^2).
		\label{kaf}
		\eeq
		We now write \eqref{op} to first order in $\hbar$ as
		\beq
		G(p^2)&=&\frac{Z}{(p^2+m_{\rm pole}^2-\widetilde{\Pi}(p^2))Z}\nonumber\\
		&=&\frac{Z}{p^2+m_{\rm pole}^2-\widetilde{\Pi}(p^2)+(p^2+m_{\rm pole}^2)\frac{\partial \widetilde{\Pi}(p^2)}{\partial p^2}\vert_{p^2=-m^2}}\nonumber\\
		&=&\frac{Z}{p^2+m_{\rm pole}^2}+Z\left(\frac{\widetilde{\Pi}(p^2)-(p^2+m_{\rm pole}^2)\frac{\partial \widetilde{\Pi}(p^2)}{\partial p^2}\vert_{p^2=-m^2}}{(p^2+m_{\rm pole}^2)^2}\right),
		\label{12}
		\eeq
		where in the last line we used a first-order Taylor expansion so that the propagator has an isolated pole at $p^2=-m_{\rm pole}^2$. In  \eqref{ltt} we can isolate this pole in the same way, by defining the spectral density function as  $\rho(t)=Z \delta(t-m_{\rm pole}^2)+\widetilde{\rho}(t)$, giving
		\beq
		G(p^2)=\frac{Z}{p^2+m_{\rm pole}^2}+ \int_0^{\infty} dt\frac{\widetilde{\rho}(t)}{t+p^2}
		\label{22}
		\eeq
		and we identify the second term in each of the representations \eqref{12} and \eqref{22} as the $\textit{reduced propagator}$
		\beq
		\widetilde{G}(p^2)&\equiv& G(p^2)-\frac{Z}{p^2+m_{\rm pole}^2},
		\eeq
		so that
		\beq
		\widetilde{G}(p^2)= \int_0^{\infty}dt \frac{\widetilde{\rho}(t)}{t+p^2} &=&Z\left(\frac{\widetilde{\Pi}(p^2)-(p^2+m_{\rm pole}^2)\frac{\partial \widetilde{\Pi}(p^2)}{\partial p^2}\vert_{p^2=-m^2}}{(p^2+m_{\rm pole}^2)^2}\right).
		\label{pp}
		\eeq
		Finally, using Cauchy's integral theorem in complex analysis, we can find the spectral density $\widetilde{\rho}(t)$ as a function of $\widetilde{G}(p^2)$, giving
		\beq
		\widetilde{\rho}(t)=\frac{1}{2\pi i}\lim_{\epsilon\to 0^+}\left(\widetilde{G}(-t-i\epsilon)-\widetilde{G}(-t+i\epsilon)\right).
		\label{key}
		\eeq
		
  Although we restricted our analysis to first order in $\hbar$ in this paper, it should not come as a surprise the foregoing methodology can be adapted order per order in $\hbar$.

		\subsection{Spectral properties of the elementary fields \label{IV}}
		We first discuss the spectral properties of the elementary fields: the transverse photon field $A^T_{\m}(x)$ and the scalar Higgs field $h(x)$.   For illustrational purposes, for the rest of this section and the next, we shall write all quantities as a function of the renormalization scale $\mu$ and choose the parameters $e=1$, $v=1 \, \mu$, $\lambda=\frac{1}{5}$, so that $m=1\, \mu$ and $m_h=\frac{1}{\sqrt{5}}\, \mu$. For this choice of parameters, all one-loop corrections computed are within $20\%$ of the tree-level results, indicating that our perturbative approximation is under control.
		
		\subsubsection{The transverse photon field \label{2b}}
		
		Since in the Abelian case the transverse component of the gauge field $A^T_{\m}(x)$ is   explicitly gauge-invariant, the corresponding propagator \eqref{dk} is independent from the gauge parameter $\xi$, and so are its pole mass, residue and spectral function.  Following the steps  from section \ref{2a}, we find the first-order pole mass of the transverse photon to be
		\beq
		m_{\rm pole}^2 &=& 1.05417\, \mu^2
		\label{polem}
		\eeq
		and the first-order residue
		\beq
		Z= 0.984983.
		\eeq
		These values are small corrections of the tree-level ones, $m^2=\mu^2$ and $Z_{\text{tree}}=1$, so that the one-loop approximation appears to be consistent.
		
\begin{figure}[t]
			\center
			\includegraphics[width=12cm]{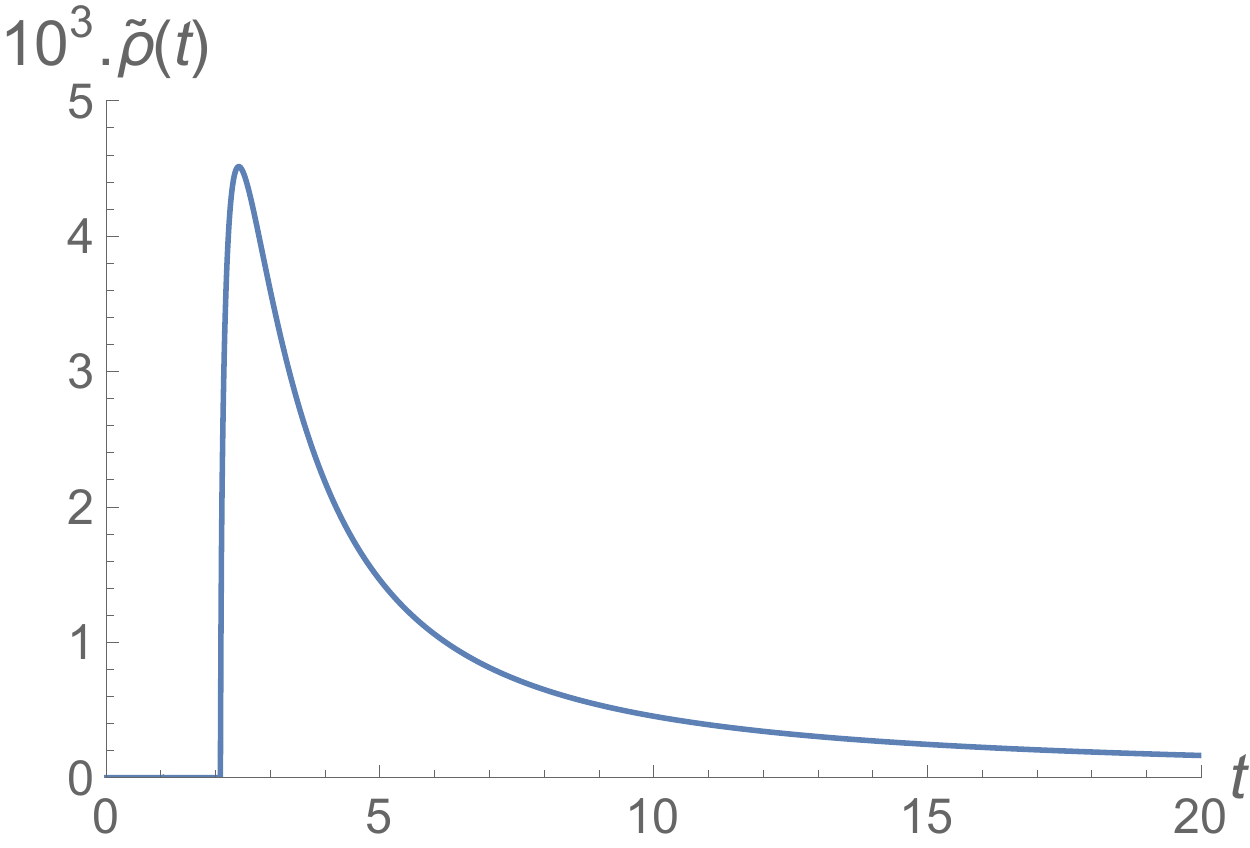}
			\caption{Spectral function for the transverse part of the reduced photon propagator $\langle A(p)A(-p) \rangle^T$, with all quantities given in units of appropriate powers of the energy scale $\mu$, for the parameter values $e=1$, $v=1 \, \mu$, $\lambda=\frac{1}{5}$.}
			\label{Y4}
		\end{figure}

		The spectral function is given in FIG.~\ref{Y4}. We can distinguish a two-particle state threshold at $t=(m+m_h)^2=2.09 \, \mu^2$, and the spectral density function is positive, adequately describing the physical photon excitation.
		
		\subsubsection{The Higgs field}
		
\begin{figure}[t]
			\center
			\includegraphics[width=12cm]{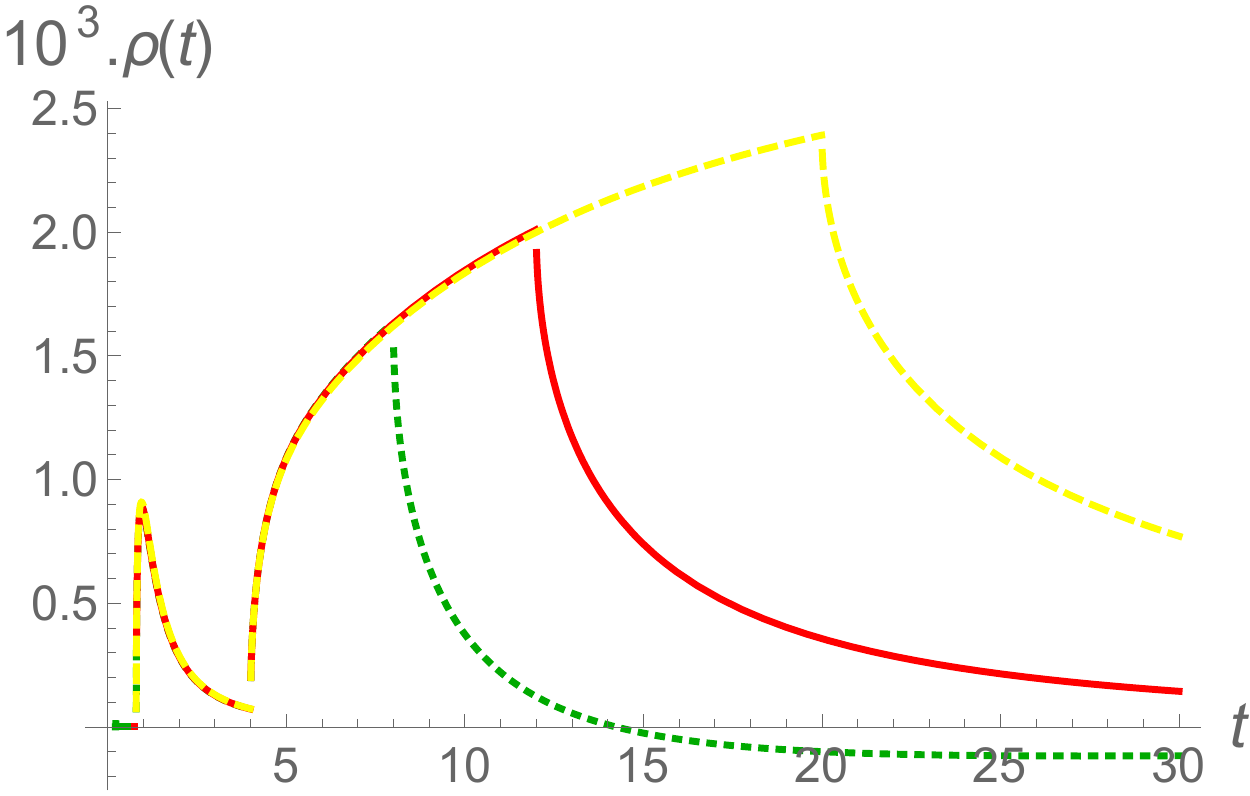}
			\caption{Spectral function for the reduced Higgs propagator $\langle h(p)h(-p) \rangle$, for gauge parameters $\xi = 2$ (Green, dotted), $\xi = 3$ (Yellow, dashed),
				$\xi= 5$ (Red, Solid),  with all quantities given in units of appropriate powers of the energy scale $\mu$, for the parameter values $e=1$, $v=1 \, \mu$, $\lambda=\frac{1}{5}$.}
			\label{Y3}
		\end{figure}

		For the Higgs fields, following the steps from section \ref{2a}, we find the pole mass to first order in $\hbar$ to be
				\beq
		m_{h,{\rm pole}}^2 &=& 0.237987 \, \mu^2=1.1899\, m_h^2,
		\label{higgsmass}
		\eeq
		for all values of the parameter $\xi$. This means that while the Higgs propagator \eqref{dk2}   is itself gauge dependent, the pole mass is gauge independent. This is in full agreement  with the Nielsen identities of the Abelian $U(1)$ Higgs model studied in \cite{Haussling:1996rq}. For the residue, we distinguish three regions:
				\begin{itemize}
			\item $\xi< \frac{1}{20}=\frac{\lambda}{4e^2}$: for these values $m_h > 2 \sqrt{\xi} m $, which means the Higgs particle is unstable and can decay into two Goldstone   modes. Of course, this process is physically impossible because the Goldstone boson itself is not physical. It therefore clearly demonstrates the unphysical nature of the propagator $\braket{h(x)h(y)}$. For these values of $\xi$, the pole mass is a   real number located on the (unphysical) branch cut created by the two-particle Goldstone state. This  means that we cannot even properly define the derivative of the one-loop correction to obtain the corresponding residue \eqref{kaf}.
			\item $\xi \leq 3$: for these values we find $Z > 1$.
			\item $\xi >3$: for these values we find $Z<1$.
		\end{itemize}
		In FIG.~\ref{Y3}, we   display the spectral density functions for three values of $\xi:2, 3, 5$. For small $t$, their behaviour is the same, with a two-particle   Higgs state at $t=(m_h+m_h)^2= 0.8\, \mu^2$, and a two-particle state for the photon field, starting at $t=(m+m)^2=4 \mu^2$. Then, we see that there is a negative contribution, different for each   case, at $t=(\sqrt{\xi} m+\sqrt{\xi} m)^2$. This corresponds to the threshold for creation of two (unphysical) Goldstone bosons. This negative contribution eventually overcomes the other ones, leading to a negative regime in the spectral function, independently of the value of $\xi$. This feature is consistent with the large-momentum behaviour of the Higgs propagator \eqref{dk2}, for a detailed discussion see \cite{us}. 		As one lowers the value of the gauge parameter $\xi$, this unphysical threshold is shifted towards lower $t$'s
		and may occur for momentum values lower than the physical two-particle states of two Higgs   particles or two photons. As discussed above, for $\xi< \frac{\lambda}{4e^2}$ even the one-particle delta peak becomes located within the unphysical Goldstone production region and the standard interpretation of the spectral properties is completely lost. It is therefore clear that this correlation function does not display the desired spectral properties to describe the Higgs mode in this theory, indicating the necessity of resorting to another operator as we shall do in what follows.

		\subsection{Spectral properties of the gauge-invariant composite operators $V_{\mu}(x)$ and $O(x)$  \label{2c}}
		
		\subsubsection{The scalar composite operator $O(x)$}
\begin{figure}[t]
			\center
			\includegraphics[width=12cm]{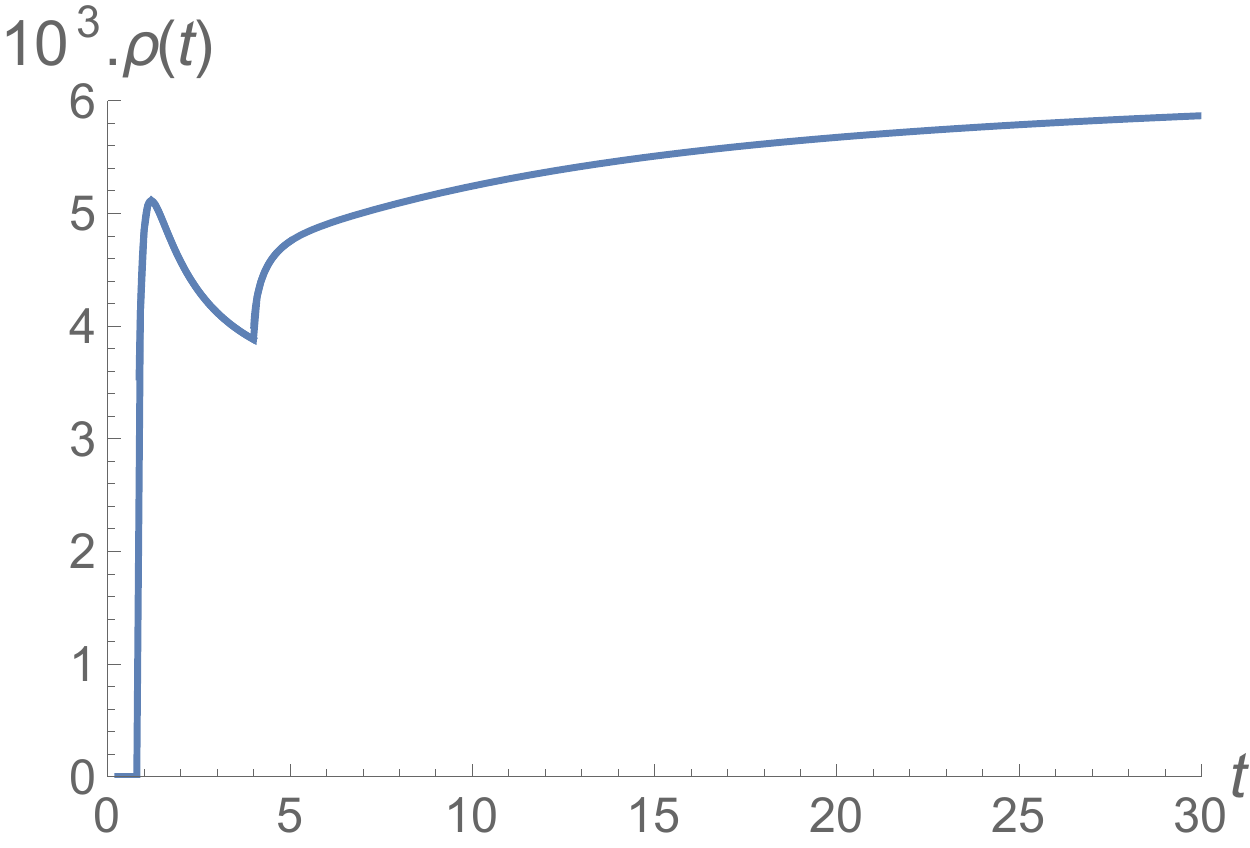}
			\caption{Spectral function for the reduced propagator of the scalar composite operator, $\langle {O}(p){O}(-p) \rangle$, with all quantities given in units of appropriate powers of the energy scale $\mu$, for the parameter values $e=1$, $v=1 \, \mu$, $\lambda=\frac{1}{5}$.}
			\label{Y}
		\end{figure}

		For the scalar composite operator ${O}(x)$ with two-point function given by expression \eqref{dk3}, we find the first-order pole mass for our set of parameter values to be
		\beq
		m^2_{h,{\rm pole}}=0.213472\, \mu^2,
		\label{rar}
		\eeq
		which is exactly equal to the pole mass of the elementary Higgs field correlator. Following the steps from \ref{2a}, we find the first-order residue to correct the tree-level result $Z_{\text{tree}}=v^2$ by $\sim 7\%$:
		\beq
		Z=v^2(1+\partial_{p^2}\Pi_{OO}(p^2)_{p^2=-m_h^2})=1.06577 v^2\,,
		\eeq
		while the first-order spectral function is shown in FIG.~\ref{Y}. Similarly as for the spectral function of the Higgs field in FIG.~\ref{Y3}, one finds a two-particle threshold for Higgs pair production at $t=(m_h+m_h)^2= 0.8\, \mu^2$, and a two-photon state starting at $t=(m+m)^2=4 \,\mu^2$. The difference is that for this gauge-invariant correlation function we no longer have the unphysical Goldstone two-particle state. Due to the absence of this negative contribution, the spectral function is always positive. Therefore, this quantity is suitable for describing a physical Higgs excitation spectrum as opposed to the elementary propagator $\langle hh\rangle$.
	
		Finally, it is interesting to note that below the unphysical threshold the elementary correlator displays the same qualitative spectral properties as this gauge-invariant approach. This means that spectral description of the physical Higgs mode could in principle be successfully encoded in the elementary propagator in the unitary gauge, in which $\xi\to \infty $ and the Goldstone bosons are infinitely heavy. We shall make an explicit comparison in section \ref{unitarylimit}.
		
		\subsubsection{The Vector composite operator $V_{\mu}(x)$}
		
\begin{figure}[t]
			\center
			\includegraphics[width=10cm]{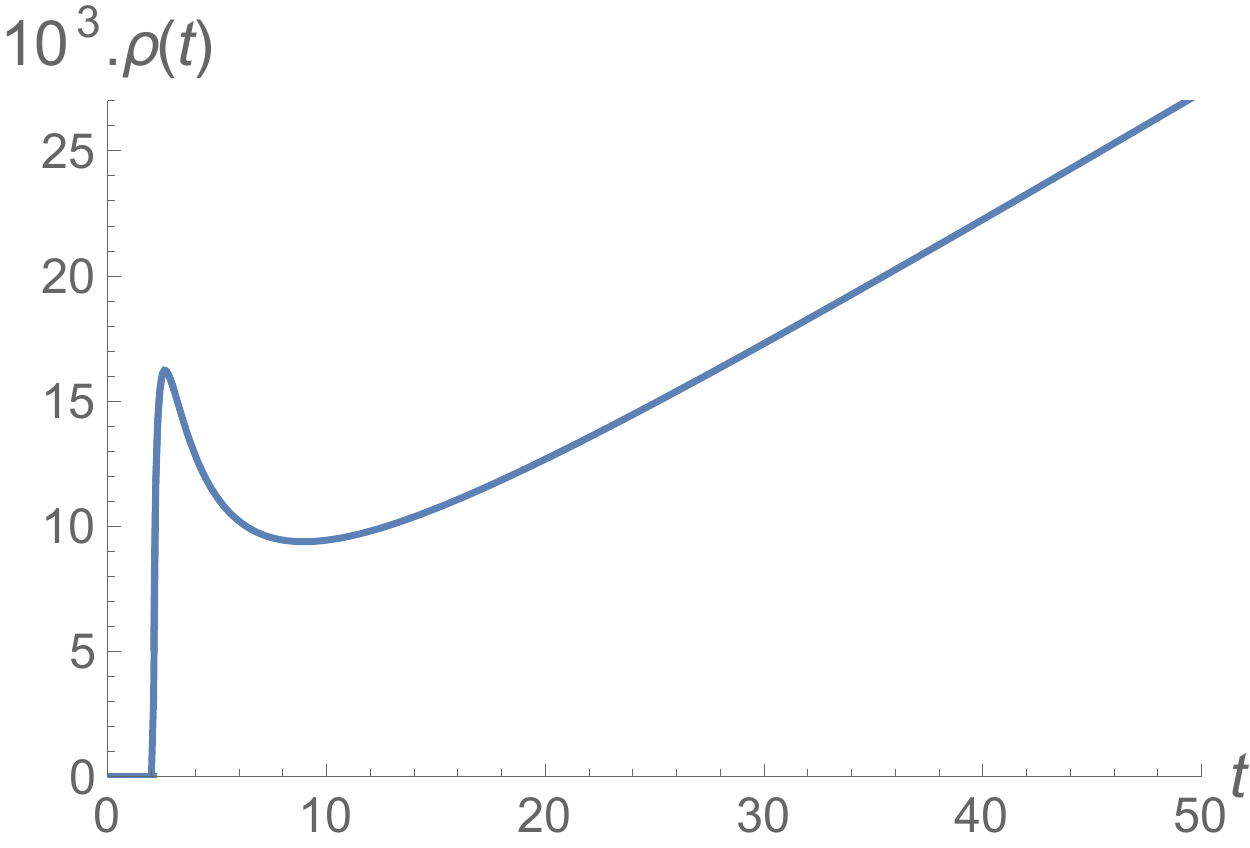}
			\caption{Spectral function for reduced transverse propagator of the vector composite operator $\langle {V}(p){V}(-p) \rangle^T$, with all quantities given in units of appropriate powers of the energy scale $\mu$, for the parameter values $e=1$, $v=1 \, \mu$, $\lambda=\frac{1}{5}$.}
			\label{W}
		\end{figure}

		For the transverse vector composite operator $V^T_{\m}(x)$, with our set of parameters we find the first-order pole mass
		\beq
		m^2_{\rm pole}=1.05417 \mu^2,
		\eeq
		which is ---as expected from the Nielsen identities--- exactly the same as the pole mass of the transverse photon field correlator \eqref{polem}. Furthermore, we find the first-order residue
		\beq
		Z=\frac{e^2v^4}{4}(1+\partial_{p^2}\Pi_{VV}^T(p^2)_{p^2=-m^2})=1.09332 \frac{e^2v^4}{4}\,,
		\eeq
		and the first order spectral density for the reduced propagator is displayed in FIG.~\ref{W}. Like the photon spectral density in FIG.~\ref{Y4}, we find a photon-Higgs two-particle state at $t=(m_h+m)^2=2.09 \, \mu^2$, and the spectral density is   again positive for all values of $t$.
		\section{  Unitary gauge limit \label{unitarylimit}}
\begin{figure}[t]
			\center
			\includegraphics[width=15cm]{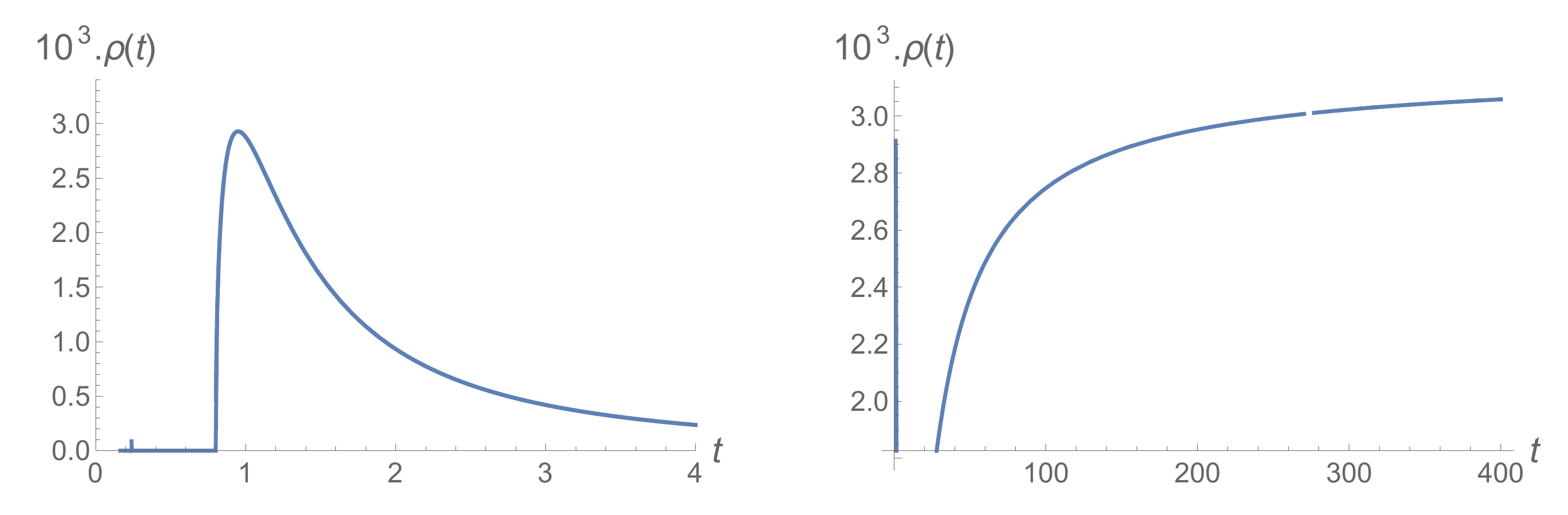}
			\caption{Spectral function for the reduced elementary propagator $\langle h(p) h(-p) \rangle$ for small values of $t$ (left) and large values of $t$ (right), with all quantities given in units of appropriate powers of the energy scale $\mu$, for $\xi=1000$ the parameter values $e=1$, $v=1 \, \mu$, $\lambda=\frac{1}{5}$.}
			\label{W3}
		\end{figure}

\begin{figure}[t]
			\center
			\includegraphics[width=15cm]{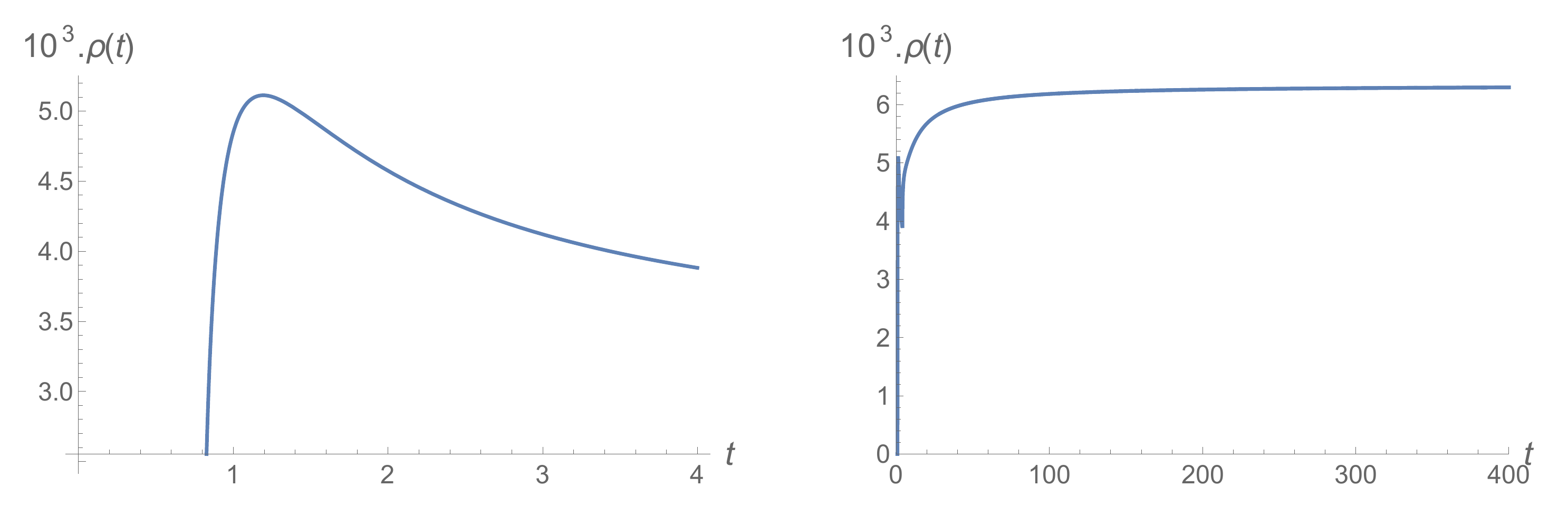}
			\caption{Spectral function for the reduced composite propagator $\langle O(p)O(-p) \rangle$ for small values of $t$ (left) and large values of $t$ (right), with all quantities given in units of appropriate powers of the energy scale $\mu$, for the parameter values $e=1$, $v=1 \, \mu$, $\lambda=\frac{1}{5}$.}
			\label{hkg}
		\end{figure}

		It is well-known \cite{Peskin:1995ev} that for the Higgs model, the unitary gauge represents the   ``most physical'' gauge, as it decouples the unphysical fields, i.e. the ghost field and the Goldstone field. The unitary gauge can be formally obtained   from  the $R_{\xi}$-gauges by taking $\xi \rightarrow \infty$. However, this gauge is non-renormalizable, as one can see by looking at this limit for the tree-level propagator of the photon field
		\beq
		\langle A_{\m}(p)A_{\n}(-p)\rangle_{\text{tree}} &\overset{\xi \rightarrow \infty}{=}& \frac{1}{p^2+ m^2}{\mathcal{P}}_{\m\n}+\frac{1}{m^2}\mathcal{L}_{\m\n}.
		\eeq
Nonetheless, we can approximate the unitary gauge by taking   larger and larger values of $\xi$. This is especially interesting when looking at the spectral function of the elementary Higgs field, which is $\xi$-dependent. In FIG.~\ref{W3} one finds the spectral function for $\xi= 1000$ for small and large ranges of $t$. In FIG.~\ref{hkg} we show the spectral function of the scalar composite field $O(x)$ for the same ranges of $t$. As one can see, the pictures are qualitatively very similar. This means that when approximating the unitary gauge, the spectral function of the gauge dependent, elementary field $h(x)$ approximates that of its composite, gauge-invariant counterpart,   thereby clearly showing the physical nature of this gauge. It is intuitively clear why this happens: all unphysical threshold effects related to ghost and Goldstone modes are pushed to higher and higher energy scale as the gauge parameter $\xi$ grows.


		\section{Conclusion and outlook \label{VI}}
		
		In the present work, following the local gauge-invariant setup of \cite{hooft1980we,hooft2012nonperturbative,Frohlich:1980gj,Frohlich:1981yi}, we have evaluated at one-loop order the two-point correlation functions $\langle V'_{\mu}(x) V'_{\nu}(y) \rangle$, $\langle O(x) O(y) \rangle$ of the two local gauge-invariant operators ${V}'_{\mu}(x)= -i \varphi^{\dagger}(x) D_{\mu} \varphi(x)+i \partial_{\m} O(x)$ and ${O}(x)= \varphi^{\dagger}(x) \varphi(x)-\frac{v^2}{2}$ in the $U(1)$ Abelian Higgs model quantized in the $R_\xi$ gauge. 

Our results can be summarized as follows:
		\begin{itemize}
			\item both $\langle V'_{\mu}(x) V'_{\nu}(y) \rangle$ and  $\langle O(x) O(y) \rangle$ do not depend on the gauge parameter $\xi$, as expected;
			\item the pole masses of $\langle V'_{\mu}(x) V'_{\nu}(y) \rangle^T$ and  $\langle O(x) O(y) \rangle$ are exactly the same as those of the correlation functions of the elementary fields $\langle A_\mu(x) A_\nu(y) \rangle^{T}$ and $\langle h(x) h(y) \rangle$, respectively, where $\langle \cdots \rangle^{T}$ stands for the transverse component of the corresponding propagator;
			\item the    K\"all\'{e}n-Lehmann spectral densities of the correlation functions $\langle V'_{\mu}(x) V'_{\nu}(y) \rangle$ and  $\langle O(x) O(y) \rangle$ turn out to be always positive, in contrast to the one associated with the (gauge dependent) elementary Higgs field.
			
		\end{itemize}
		These important features give us a fully gauge-invariant picture in order to describe the spectrum of elementary excitations of the model, i.e.~the massive photon and the Higgs mode. 

It is worth underlining that the local gauge-invariant operators $V'_{\mu}(x)$ and $O(x)$ have their generalization to the non-Abelian case \cite{hooft1980we,hooft2012nonperturbative,Frohlich:1980gj,Frohlich:1981yi}. \footnote{See also the recent work \cite{Sondenheimer:2019idq2} for a discussion  on higher dimensional gauge-invariant operators for different gauge groups and representations of the Higgs fields.}This might enable us to extend the present work to the case of asymptotically free Higgs-Yang-Mills theories such as, for example, the $SU(2)$ theory with a single Higgs field in the fundamental representation, see for example \cite{Kondo:2018qus}. This model might be of particular interest since non-perturbative   effects can be introduced in order to achieve a better understanding of   $SU(2)$ Higgs-Yang-Mills theory and the fate of the excitations in the infrared region.   Indeed, as there is not even an strict order parameter discriminating between confinement- or Higgs-like behaviour in such theory, it should intuitively be possible to interpolate from one behaviour to the other without encountering sharp phase boundaries, a feature potentially encoded in the gauge-invariant correlation functions.

 More specifically, one may for example introduce the   Gribov-Zwanziger horizon term, in its BRST-invariant formulation encoded in the so called Refined Zwanziger-Gribov action (cf. \cite{Vandersickel:2012tz,Dudal:2008sp,Capri:2015ixa,Capri:2018ijg} and refs. therein) implementing the restriction to the Gribov region $\Omega$ \cite{gribov1978quantization} in order to take into account the existence of the Gribov copies plaguing the   non-Abelian Faddeev-Popov quantization procedure.
		As a consequence, the gauge-invariant pole masses of the non-Abelian generalization of the correlation functions $\langle V_{\mu}(x) V_{\nu}(y) \rangle$ and  $\langle O(x) O(y) \rangle$ will now show an explicit dependence on the   (BRST invariant) Gribov mass parameter as well as on the dimension-two condensates present in the Refined-Gribov-Zwanziger action \cite{Vandersickel:2012tz,Dudal:2008sp,Capri:2015ixa,Capri:2018ijg}. Thus, extending the framework already outlined in \cite{Capri:2012ah}, the aforementioned  pole masses and further spectral properties could be employed as gauge-invariant probing quantities in order to extract non-perturbative information about the behaviour of the excitations of Higgs-Yang-Mills theories in the light of the Fradkin-Shenker \cite{Fradkin:1978dv,Caudy:2007sf} results. 

  Another most interesting extension of our methodology would be to the Glashow-Weinberg-Salam electroweak theory, to have a genuinely gauge-invariant description of at least the $W^\pm,Z^0$- and Higgs boson sector of the theory, including their spectral functions bearing information on both pole mass and decay channels \cite{Jegerlehner:2001fb,Jegerlehner:2002em,Martin:2015lxa,Martin:2015rea}.

  We hope to report soon on these interesting and relevant issues.

		\section*{Acknowledgements}
		The authors would like to thank the Brazilian agencies CNPq and FAPERJ for financial support. This study was financed in part by the Coordenação de Aperfeiçoamento de Pessoal de Nível Superior---Brasil (CAPES)--- Finance Code~001. This paper is also part of the project INCT-FNA Process No.~464898/2014-5.

		\appendix

		\section{Propagators and vertices of the Abelian Higgs model in the $R_{\xi}$ gauge\label{FR}}

		\subsection{Field propagators}
		The quadratic part of the action \eqref{fullaction} in the bosonic sector is given by
		\beq
		S_{\rm bos}^{\rm quad}&=&\ha\int d^4 x \Big\{A_{\m}(-\d_{\m\n}(\pa^2-m^2)+\pa_{\m}\pa_{\n})A_{\n}-\r \pa^2 \r  -h(\pa^2 - m_{h}^2)h +\bar{c}(\pa^2 -m^2 \xi)c\nonumber\\
		&+&2ib\pa_{\m}A_{\m}+\xi b^2+2im\xi b  \rho +2m A_{\m}\pa_{\m}\rho \Big\}.
		\eeq
		Putting this in a matrix form yields
		\beq
		S^{quad}_{bos}=\ha\int d^4 x \,\Psi^{T}_{\m} {O}_{\m\n} \Psi_{\n},
		\eeq
		where
		\beq
		\Psi^{T}_{\m}=\left( {\begin{array}{cccc}
				A_{\m} &
				b&
				\r&
				h
		\end{array} } \right),\,\, \Psi_{\n}=\left( {\begin{array}{cccc}
				A_{\n}\nonumber \\
				b\nonumber \\
				\r\nonumber \\
				h
		\end{array} } \right),
		\eeq
		and
		\beq
		  O_{\mu\nu}=\left( {\begin{array}{cccc}
				(-\d_{\m\n}(\pa^2-m^2)+\pa_{\m}\pa_{\n})&-i\pa_{\mu}&m\pa_{\m}&0\nonumber \\
				i \pa_{\n} &\xi &i m \xi &0\nonumber \\
				-m \partial_{\n} &i m \xi&-\pa^2&0\nonumber \\
				0&0&0&-(\pa^2 - m_{h}^2)
		\end{array} } \right).
		\eeq
		The tree-level field propagators can be read off from the inverse of $\mathcal{O}$, leading to the following expressions   in momentum space:
		\beq
		\langle A_{\m}(p)A_{\n}(-p)\rangle &=& \frac{1}{p^2+ m^2}{\mathcal{P}}_{\m\n}+\frac{\xi}{p^2 + \xi m^2}\mathcal{L}_{\m\n},\nonumber\nonumber \\
		\langle \r(p)\r(-p)\rangle &=&\frac{1}{p^2 +\xi m^2},\nonumber\\
		\langle h(p)h(-p)\rangle &=&\frac{1}{p^2 + m_h^2},\nonumber\\
		\langle A_{\m}(p)b(-p)\rangle &=& \frac{p_{\m}}{p^2+\xi m^2},\nonumber\\
		\langle b (p) \rho (-k) \rangle &=& \frac{-i m}{p^2+ \xi m^2},
		\eeq
		where $\mathcal{P}_{\m\n}=\delta_{\m\n}-\frac{p_{\m}p_{\n}}{p^2}$ and $\mathcal{L}_{\m\n}=\frac{p_{\m}p_{\n}}{p^2}$ are the transversal and longitudinal projectors, respectively. The ghost propagator is
		\beq
		\langle \bar{c}(p)c(-p)\rangle &=&\frac{1}{p^2 +\xi m^2}.
		\eeq
		
		\subsection{Vertices}
		From the action \eqref{fullaction}, we find the following vertices
		\beq
		\Gamma_{A_{\m}\r h}(-p_1,-p_2,-p_3)&=&ie(p_{\m,3}-p_{\m,2})\delta(p_1+p_2+p_3),\nonumber\\
		\Gamma_{A_{\m}A_{\n}h}(-p_1,-p_2,-p_3)&=& -2e^2 v \delta_{\m\n}\delta(p_1+p_2+p_3),\nonumber\\
		\Gamma_{A_{\m}A_{\n}hh}(-p_1,-p_2,-p_3,-p_4)&=&-2e^2 \delta_{\m\n}\delta(p_1+p_2+p_3+p_4),\nonumber\\
		\Gamma_{A_{\m}A_{\n}\r\r}(-p_1,-p_2,-p_3,-p_4)&=&-2e^2 \delta_{\m\n}\delta(p_1+p_2+p_3+p_4),\nonumber\\
		\Gamma_{hhhh}(-p_1,-p_2,-p_3,-p_4)&=&-3\l \, \delta(p_1+p_2+p_3+p_4),\nonumber \\
		\Gamma_{hh\r \r}(-p_1,-p_2,-p_3,-p_4)&=&-\l \, \delta(p_1+p_2+p_3+p_4), \nonumber\\
		\Gamma_{\r\r\r\r}(-p_1,-p_2,-p_3,-p_4)&=&-3\l \,\delta(p_1+p_2+p_3+p_4), \nonumber\\
		\Gamma_{hhh}(-p_1,-p_2,-p_3)&=&-3\l v \,\delta(p_1+p_2+p_3),\nonumber \\
		\Gamma_{h\r\r}(-p_1,-p_2,-p_3)&=&-\l v \,\delta(p_1+p_2+p_3), \nonumber\\
		\Gamma_{\bar{c}h c}(-p_1,-p_2,-p_3)&=&-m\xi e  \,\delta(p_1+p_2+p_3).
		\eeq
		\section{  Basic Feynman integral \label{app}}
		\begin{multline}
		\int_{0}^{1}dx\;\ln\left[\frac{p^{2}x\left(1-x\right)+xm_{1}^{2}+\left(1-x\right)m_{2}^{2}}{\mu^{2}}\right]=\frac{1}{2p^{2}}\left\{ \left(m_{1}^{2}-m_{2}^{2}\right)\ln\left(\frac{m_{2}^{2}}{m_{1}^{2}}\right)+p^{2}\ln\left(\frac{m_{1}^{2}m_{2}^{2}}{\mu^{4}}\right)-4p^{2}\right.\\
		-2i\sqrt{\left(m_{1}^{2}-m_{2}^{2}\right)^{2}+p^{4}+2p^{2}\left(m_{1}^{2}+m_{2}^{2}\right)}\left[\tan^{-1}\left(\frac{-m_{1}^{2}+m_{2}^{2}-p^{2}}{\sqrt{-m_{1}^{4}+2m_{1}^{2}\left(m_{2}^{2}-p^{2}\right)-\left(m_{2}^{2}+p^{2}\right)^{2}}}\right)\right.\\
		\left.\left.-\tan^{-1}\left(\frac{-m_{1}^{2}+m_{2}^{2}+p^{2}}{\sqrt{-m_{1}^{4}+2m_{1}^{2}\left(m_{2}^{2}-p^{2}\right)-\left(m_{2}^{2}+p^{2}\right)^{2}}}\right)\right]\right\} 
		\end{multline}
		\section{Contributions to $\langle O(p) O(-p) \rangle$ \label{A}}
		\begin{figure}[t]
\centering
			\includegraphics[width=15cm]{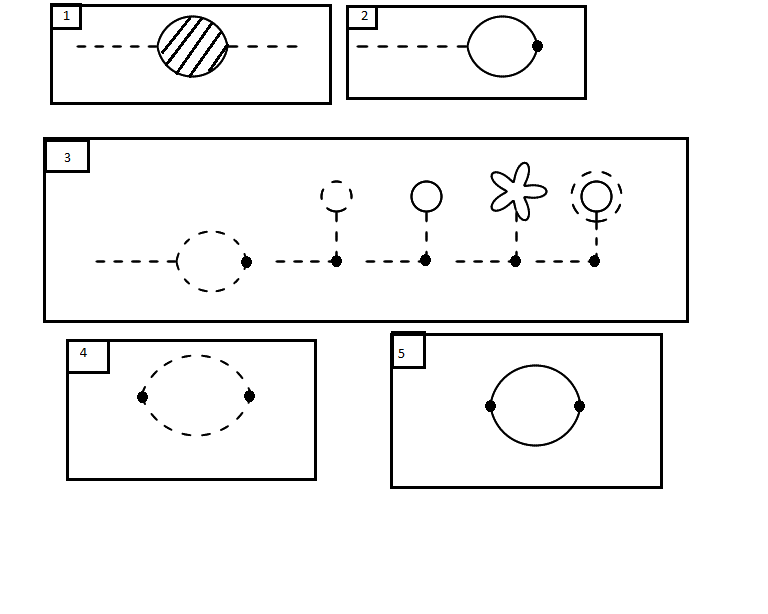}
			\caption{One-loop contributions to the propagator $\langle O(p)O(-p) \rangle$. Wavy lines represent the photon field, dashed lines the Higgs field, solid lines the Goldstone boson and double lines the ghost field.   The $\bullet$ indicates the insertion of a composite operator.}
						\label{Yw}
		\end{figure}
		We consider each term in the two-point function $\braket{O(p)O(-p)}$, given by eq. \eqref{exp1}. We will use the following definitions:
		
		\beq
		\eta(m_1,m_2,p^2)&\equiv& \frac{1}{(4\pi)^{d/2}}\Gamma(2-\frac{d}{2})\int_0^1 dx \left(p^2 x(1-x)+x m_1+(1-x) m_2\right)^{d/2-2}\nonumber\\
		\chi(m_1)&\equiv&  \frac{1}{(4\pi)^{d/2}} \Gamma(1-\frac{d}{2}) m_1^{d/2-1}.
		\eeq
		The first term is the one-loop correction to the Higgs propagator $\langle h(p)h(-p) \rangle$ known from \cite{us}, shown in frame $\fbox{1}$ in FIG.~\ref{Yw}, which gives
			\begin{eqnarray}
			v\braket{ h\left(p\right)\rho\left(-p\right)^{2}}  & = & -\frac{m_{h}^{2}}{p^{2}+m_{h}^{2}}\eta\left(\xi m^{2},\xi m^{2},p^{2}\right)
			\end{eqnarray}
			
		The second term, the one-loop correction shown in frame $\fbox{2}$ of FIG.~\ref{Yw}, gives
	\begin{eqnarray}
	v\langle h\left(p\right)\rho\left(-p\right)^{2}\rangle  & = & -\frac{m_{h}^{2}}{p^{2}+m_{h}^{2}}\eta\left(\xi m^{2},\xi m^{2},p^{2}\right)
	\end{eqnarray}
	
		The third term, the one-loop correction shown in frame  $\fbox{3}$ of FIG.~\ref{Yw}, gives
				\begin{eqnarray}
				v\langle h\left(p\right)h\left(-p\right)^{2}\rangle  & = & \frac{1}{p^{2}+m_{h}^{2}}\left\{ -3m_{h}^{2}\eta\left(m_{h}^{2},m_{h}^{2},p^{2}\right)-3\chi\left(m_{h}^{2}\right)-\frac{2\left(d-1\right)m^{2}}{m_{h}^{2}}\chi\left(m^{2}\right)-\chi\left(\xi m^{2}\right)\right\} 
				\end{eqnarray}
				
		The fourth term has no one-loop contributions.
		The fifth term, the one-loop correction shown in frame $\fbox{4}$ of FIG.~\ref{Yw}, gives
	\begin{eqnarray}
	\frac{1}{4}\langle h\left(p\right)^{2}h\left(-p\right)^{2}\rangle  & = & \frac{1}{2}\eta\left(m_{h}^{2},m_{h}^{2},p^{2}\right).
	\end{eqnarray}
		The sixth term, the one-loop correction shown in frame $\fbox{5}$ of FIG.~\ref{Yw}, gives
	
	\begin{eqnarray}
	\frac{1}{4}\langle \rho\left(p\right)^{2}\rho\left(-p\right)^{2}\rangle  & = & \frac{1}{2}\eta\left(\xi m^{2},\xi m^{2},p^{2}\right)
	\end{eqnarray}
		Using the identity \eqref{2o} we are able to write the whole  one-loop correlation
function $\braket{O(-p) O(p)}$, up to the order $\hbar$,  as
\begin{eqnarray}
\langle O\left(p\right)O\left(-p\right)\rangle  & = & \frac{v^{2}}{p^{2}+m_{h}^{2}}+\frac{1}{\left(p^{2}+m_{h}^{2}\right)^{2}}\left\{ \frac{1}{2}\left[4\left(d-1\right)m^{4}+4m^{2}p^{2}+p^{4}\right]\eta\left(m^{2},m^{2},p^{2}\right)\right.\nonumber \\
&  & \left.+\frac{1}{2}\left(p^{2}-2m_{h}^{2}\right)^{2}\eta\left(m_{h}^{2},m_{h}^{2},p^{2}\right)-p^{2}\left[2\left(d-1\right)\frac{m^{2}}{m_{h}^{2}}+1\right]\chi\left(m^{2}\right)-3p^{2}\chi\left(m_{h}^{2}\right)\right\} 
\label{ch}
\end{eqnarray}

		\section{Contributions to $\langle V_{\mu}(x)V_{\nu}(y)\rangle$ \label{ah}}
		\begin{figure}[t]
			\centering
			\includegraphics[width=19cm]{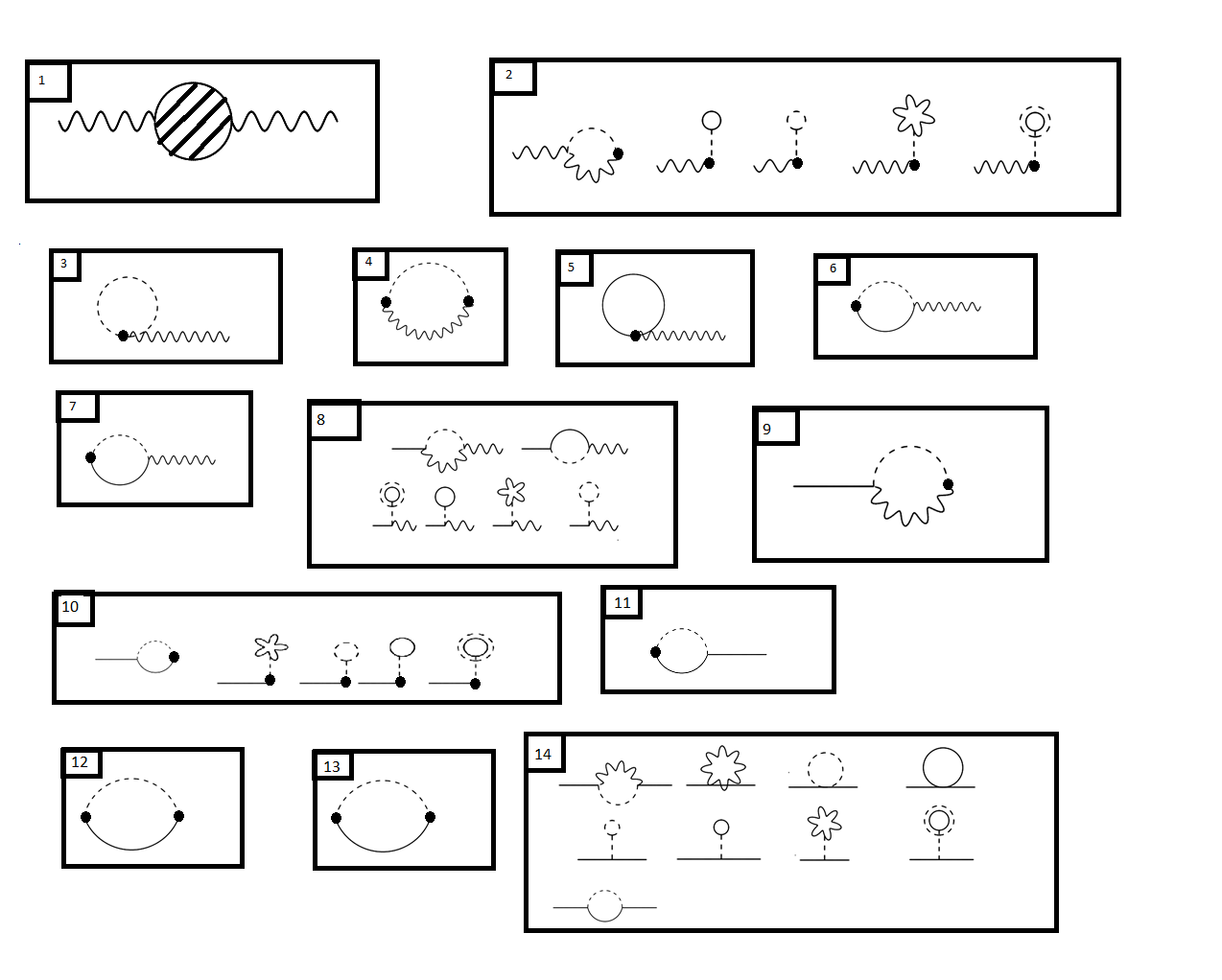}
			\caption{One-loop contributions for the propagator $\langle V_{\mu}(p)V_{\nu}(-p) \rangle$.  Wavy lines represent the photon field, dashed lines the Higgs field, solid lines the Goldstone boson and double lines the ghost field.   The $\bullet$ indicates the insertion of a composite operator.}
			\label{Y22}
		\end{figure}
		We consider each term in the two-point function $\braket{V_{\m}(p)V_{\n}(-p)}$, given by eq. \eqref{exp2}. The first term is the one-loop correction to the photon propagator $\langle A_{\m}(p)A_{\n}(-p) \rangle$ known from \cite{us}, shown in frame $\fbox{1}$ in FIG.~\ref{Y22}, which gives
	\begin{eqnarray}
	\frac{e^{2}v^{4}}{4}\langle A_{\mu}\left(p\right)A_{\nu}\left(-p\right)\rangle_{1-{\rm loop}}  & = & \frac{e^{2}v^{4}}{4}\mathcal{P}_{\mu\nu}\left(p\right)\nonumber  \Bigg\{\frac{1}{\left(p^{2}+m^{2}\right)^{2}}\left[-\frac{m^{2}\left(\left(m_{h}^{2}-m^{2}+p^{2}\right)^{2}-4\left(d-2\right)m^{2}p^{2}\right)}{\left(d-1\right)v^{2}p^{2}}\eta\left(m^{2},m_{h}^{2},p^{2}\right)\right.\nonumber \\
	&  & +\frac{m^{2}\left(2\left(d-1\right)^{2}m^{2}p^{2}+m_{h}^{2}\left(p^{2}-m^{2}\right)+m_{h}^{4}\right)}{\left(d-1\right)v^{2}p^{2}m_{h}^{2}}\chi\left(m^{2}\right)\nonumber \\
	&  & \left.\left.+\frac{m^{2}\left(\left(2d-1\right)p^{2}-m_{h}^{2}+m^{2}\right)}{\left(d-1\right)v^{2}p^{2}}\chi\left(m_{h}^{2}\right)\right]\right\} \nonumber \\
	&  & +\frac{e^{2}v^{4}}{4}\mathcal{L}_{\mu\nu}\left(p\right) \nonumber  \Bigg\{\frac{\xi^{2}}{\left(p^{2}+\xi m^{2}\right)^{2}}\left[\frac{m^{2}\left(-2m_{h}^{2}\left(m^{2}-p^{2}\right)+m_{h}^{4}+\left(m^{2}+p^{2}\right)^{2}\right)}{v^{2}p^{2}}\eta\left(m^{2},m_{h}^{2},p^{2}\right)\right.\nonumber \\
	&  & -\frac{m^{2}\left(2m_{h}^{2}-2\xi m^{2}+p^{2}\right)}{v^{2}}\eta\left(m_{h}^{2},\xi m^{2},p^{2}\right)\nonumber \\
	&  & +\frac{m^{2}\left(2\left(d-1\right)m^{2}p^{2}+m_{h}^{2}\left(m^{2}-p^{2}\right)-m_{h}^{4}\right)}{v^{2}p^{2}m_{h}^{2}}\chi\left(m^{2}\right)\nonumber \\
	&  & \left.\left.-\frac{m^{2}\left(-m_{h}^{2}+m^{2}-3p^{2}\right)}{v^{2}p^{2}}\chi\left(m_{h}^{2}\right)+2\frac{m^{2}}{v^{2}}\chi\left(\xi m^{2}\right)\right]\right\} 
	\end{eqnarray}
		The second term, the one-loop correction shown in frame $\fbox{2}$ of FIG.~\ref{Y22}, gives
		\begin{eqnarray}
		e^{2}v^{3}\langle \left(hA_{\mu}\right)\left(p\right)A_{\nu}\left(-p\right)\rangle  & = & \frac{e^{2}v^{3}}{p^{2}+m^{2}}\mathcal{P}_{\mu\nu}\left(p\right)\left\{ -\frac{e\left[2evm_{h}^{4}\left(p^{2}-\xi m^{2}\right)+evm_{h}^{2}\left(p^{2}+\xi m^{2}\right)^{2}+evm_{h}^{6}\right]}{2\left(d-1\right)m^{2}m_{h}^{2}p^{2}}\eta\left(m_{h}^{2},\xi m^{2},p^{2}\right)\right.\nonumber \\
		&  & -\frac{e\left[evm_{h}^{2}\left(-2\left(3-2d\right)m^{2}p^{2}-m^{4}-p^{4}\right)+2evm_{h}^{4}\left(m^{2}-p^{2}\right)-evm_{h}^{6}\right]}{2\left(d-1\right)m^{2}m_{h}^{2}p^{2}}\eta\left(m^{2},m_{h}^{2},p^{2}\right)\nonumber \\
		&  & -\frac{e\left[\left(d-2\right)evm_{h}^{2}p^{2}+\xi evm^{2}m_{h}^{2}-evm_{h}^{4}\right]}{2\left(d-1\right)m^{2}m_{h}^{2}p^{2}}\chi\left(\xi m^{2}\right)\nonumber \\
		&  & -\frac{e\left[3devm_{h}^{2}p^{2}-\xi evm^{2}m_{h}^{2}+evm^{2}m_{h}^{2}-3evm_{h}^{2}p^{2}\right]}{2\left(d-1\right)m^{2}m_{h}^{2}p^{2}}\chi\left(m_{h}^{2}\right)\nonumber \\
		&  & \left.-\frac{e\left[2\left(d-1\right)^{2}evm^{2}p^{2}-evm^{2}m_{h}^{2}+evm_{h}^{2}p^{2}+evm_{h}^{4}\right]}{2\left(d-1\right)m^{2}m_{h}^{2}p^{2}}\chi\left(m^{2}\right)\right\} \nonumber \\
		&  & +\frac{e^{2}v^{3}\xi}{p^{2}+\xi m^{2}}\frac{1}{2vp^{2}}\mathcal{L}_{\mu\nu}\left(p\right)\Bigg\{ \left(m_{h}^{2}-\xi m^{2}+p^{2}\right)^{2}\eta\left(m_{h}^{2},\xi m^{2},p^{2}\right)\nonumber \\
		&  & -\left[-2m_{h}^{2}\left(m^{2}-p^{2}\right)+m_{h}^{4}+\left(p^{2}+m^{2}\right)^{2}\right]\eta\left(m^{2},m_{h}^{2},p^{2}\right)\nonumber \\
		&  & +\frac{\left[-2\left(d-1\right)m^{2}p^{2}+m_{h}^{2}\left(p^{2}-m^{2}\right)+m_{h}^{4}\right]}{m_{h}^{2}}\chi\left(m^{2}\right)\nonumber \\
		&  &-\left(m_{h}^{2}-\xi m^{2}+2p^{2}\right)\chi\left(\xi m^{2}\right)-\left[m^{2}\left(\xi-1\right)+3p^{2}\right]\chi\left(m_{h}^{2}\right)\Bigg\} 
		\end{eqnarray}
		The third term, the one-loop correction shown in frame $\fbox{3}$ of FIG.~\ref{Y22}, gives
	\begin{eqnarray}
	\frac{e^{2}v^{2}}{2}\langle \left(h^{2}A_{\mu}\right)\left(p\right)A_{\nu}\left(-p\right)\rangle  & = & \frac{e^{2}v^{2}}{2}\left(\frac{1}{p^{2}+m^{2}}\mathcal{P}_{\mu\nu}\left(p\right)+\frac{\xi}{p^{2}+\xi m^{2}}\mathcal{L}_{\mu\nu}\left(p\right)\right)\chi\left(m_{h}^{2}\right)
	\end{eqnarray}
	
	The fourth term, the one-loop correction shown in frame $\fbox{4}$ of FIG.~\ref{Y22}, gives
		\begin{eqnarray}
		e^{2}v^{2}\langle \left(hA_{\mu}\right)\left(p\right)\left(hA_{\nu}\right)\left(-p\right)\rangle  & = & e^{2}v^{2}\mathcal{P}_{\mu\nu}\left(p\right)\left\{ \frac{\left(m_{h}^{2}-\xi m^{2}+p^{2}\right)^{2}+4\xi m^{2}p^{2}}{4m^{2}p^{2}\left(d-1\right)}\eta\left(m_{h}^{2},\xi m^{2},p^{2}\right)\right.\nonumber \\
		&  & +\frac{4\left(d-2\right)m^{2}p^{2}-\left(m_{h}^{2}-m^{2}+p^{2}\right)^{2}}{4\left(d-1\right)m^{2}p^{2}}\eta\left(m^{2},m_{h}^{2},p^{2}\right)\nonumber \\
		&  & -\frac{\left(m_{h}^{2}-\xi m^{2}+p^{2}\right)}{4\left(d-1\right)m^{2}p^{2}}\chi\left(\xi m^{2}\right)+\frac{\left(m_{h}^{2}-m^{2}+p^{2}\right)}{4\left(d-1\right)m^{2}p^{2}}\chi\left(m^{2}\right)\nonumber \\
		&  & \left.-\frac{\left(\xi-1\right)}{4\left(d-1\right)p^{2}}\chi\left(m_{h}^{2}\right)\right\} \nonumber \\
		&  & +\frac{e^{2}v^{2}}{4m^{2}p^{2}}\mathcal{L}_{\mu\nu}\left(p\right)\left\{ -\left(m_{h}^{2}-\xi m^{2}+p^{2}\right)^{2}\eta\left(m_{h}^{2},\xi m^{2},p^{2}\right)\right.\nonumber \\
		&  & +\left[\left(m_{h}^{2}-m^{2}+p^{2}\right)^{2}+4m^{2}p^{2}\right]\eta\left(m^{2},m_{h}^{2},p^{2}\right)\nonumber \\
		&  & +m^{2}\left(\xi-1\right)\chi\left(m_{h}^{2}\right)+\left(m_{h}^{2}-\xi m^{2}+p^{2}\right)\chi\left(\xi m^{2}\right)\nonumber \\
		&  & \left.+\left(-m_{h}^{2}+m^{2}-p^{2}\right)\chi\left(m^{2}\right)\right\} 
		\end{eqnarray}
		
		The fifth term, the one-loop correction shown in frame $\fbox{5}$ of FIG.~\ref{Y22}, gives
	
	\begin{eqnarray}
	\frac{e^{2}v^{2}}{2}\langle \left(\rho^{2}A_{\mu}\right)\left(p\right)A_{\nu}\left(-p\right)\rangle  & = & \frac{e^{2}v^{2}}{2}\left(\frac{1}{p^{2}+m^{2}}\mathcal{P}_{\mu\nu}\left(p\right)+\frac{\xi}{p^{2}+\xi m^{2}}\mathcal{L}_{\mu\nu}\left(p\right)\right)\chi\left(\xi m^{2}\right)
	\end{eqnarray}
		The sixth term, the one-loop correction shown in frame $\fbox{6}$ of FIG.~\ref{Y22}, gives
	\begin{eqnarray}
	-\frac{iev^{2}}{2}p_{\mu}\langle \left(h\rho\right)\left(p\right)A_{\nu}\left(-p\right)\rangle  & = & -\frac{iev^{2}}{2}\frac{\xi}{p^{2}+\xi m^{2}}\mathcal{L}_{\mu\nu}\left(p\right)\left[ie\left(\xi m^{2}-m_{h}^{2}\right)\eta\left(m_{h}^{2},\xi m^{2},p^{2}\right)-ie\chi\left(m_{h}^{2}\right)+ie\chi\left(\xi m^{2}\right)\right]
	\end{eqnarray}
		The seventh term, the one-loop correction shown in frame $\fbox{7}$ of FIG.~\ref{Y22}, gives
	\begin{eqnarray}
	ev^{2}\langle \left(\rho\partial_{\mu}h\right)\left(p\right)A_{\nu}\left(-p\right)\rangle  & = & \frac{ev^{2}}{2\left(d-1\right)p^{2}}\frac{1}{p^{2}+m^{2}}\mathcal{P}_{\mu\nu}\left(p\right)\left\{ -e\left[\left(-m_{h}^{2}+\xi m^{2}+p^{2}\right)^{2}+4m_{h}^{2}p^{2}\right]\eta\left(m_{h}^{2},\xi m^{2},p^{2}\right)\right.\nonumber \\
	&  & \left.+e\left(m_{h}^{2}-\xi m^{2}+p^{2}\right)\chi\left(\xi m^{2}\right)+e\left(-m_{h}^{2}+\xi m^{2}+p^{2}\right)\chi\left(m_{h}^{2}\right)\right\} \nonumber \\
	&  & +\frac{ev^{2}}{2p^{2}}\frac{\xi}{p^{2}+\xi m^{2}}\mathcal{L}_{\mu\nu}\left(p\right)\left\{ e\left(m_{h}^{2}-\xi m^{2}\right)\chi\left(m_{h}^{2}\right)+e\left(-m_{h}^{2}+\xi m^{2}+2p^{2}\right)\chi\left(\xi m^{2}\right)\right.\nonumber \\
	&  & \left.+e\left[-3p^{2}\left(m_{h}^{2}-\xi m^{2}+p^{2}\right)+\left(m_{h}^{2}-\xi m^{2}+p^{2}\right)^{2}+2p^{4}\right]\eta\left(\xi m^{2},m_{h}^{2},p^{2}\right)\right\} 
	\end{eqnarray}
			The eighth term, the one-loop correction shown in frame $\fbox{8}$ of FIG.~\ref{Y22}, gives
			\begin{eqnarray}
			-\frac{iev^{3}}{2}p_{\mu}\langle \rho\left(p\right)A_{\nu}\left(-p\right)\rangle  & = & -\frac{iev^{3}}{2}\frac{\xi}{\left(p^{2}+\xi m^{2}\right)^{2}}\mathcal{L}_{\mu\nu}\left(p\right)\left\{ \frac{ie^{3}v\left[\left(m_{h}^{2}-m^{2}+p^{2}\right)^{2}+4m^{2}p^{2}\right]}{m^{2}}\eta\left(m^{2},m_{h}^{2},p^{2}\right)\right.\nonumber \\
			&  & +\frac{\left[ie^{3}vm_{h}^{2}\left(m_{h}^{2}-\xi m^{2}\right)-ie^{3}v\left(p^{2}+m_{h}^{2}\right)\left(m_{h}^{2}-\xi m^{2}+p^{2}\right)\right]}{m^{2}}\eta\left(m_{h}^{2},\xi m^{2},p^{2}\right)\nonumber \\
			&  & +\left[\frac{i\left(d-1\right)e^{3}vp^{2}}{m_{h}^{2}}-\frac{ie^{3}v\left(m_{h}^{2}-m^{2}+p^{2}\right)}{m^{2}}\right]\chi\left(m^{2}\right)\nonumber \\
			&  & \left.+\left[\frac{2ie^{3}vm_{h}^{2}+3ie^{3}vp^{2}-2ie^{3}vm^{2}}{2m^{2}}\right]\chi\left(m_{h}^{2}\right)+\frac{3ie^{3}vp^{2}}{2m^{2}}\chi\left(\xi m^{2}\right)\right\} 
			\end{eqnarray}
		The ninth term, the one-loop correction shown in frame $\fbox{9}$ of FIG.~\ref{Y22}, gives
		\begin{eqnarray}
		-iev^{3}p_{\mu}\langle \rho\left(p\right)\left(hA_{\nu}\right)\left(-p\right)\rangle  & = & \frac{1}{2\left(p^{2}+\xi m^{2}\right)}\mathcal{L}_{\mu\nu}\left(p\right)\left\{ \left(p^{2}+m_{h}^{2}\right)\left(m_{h}^{2}-\xi m^{2}+p^{2}\right)\eta\left(\xi m^{2},m_{h}^{2},p^{2}\right)\right.\nonumber \\
		&  & -\left[\left(m_{h}^{2}-m^{2}+p^{2}\right)^{2}+4m^{2}p^{2}\right]\eta\left(m^{2},m_{h}^{2},p^{2}\right)\nonumber \\
		&  & \left.-\left(p^{2}+m_{h}^{2}\right)\chi\left(\xi m^{2}\right)+m^{2}\chi\left(m_{h}^{2}\right)+\left(p^{2}+m_{h}^{2}-m^{2}\right)\chi\left(m^{2}\right)\right\} 
		\end{eqnarray}
		The tenth term, the one-loop correction shown in frame $\fbox{10}$ of FIG.~\ref{Y22}, gives
		\begin{eqnarray}
		-\frac{v}{2}p_{\mu}p_{\nu}\langle \left(h\rho\right)\left(p\right)\rho\left(-p\right)\rangle  & = & \frac{vp^{2}}{2}\frac{1}{p^{2}+\xi m^{2}}\mathcal{L}_{\mu\nu}\left(p\right)\left[\frac{e^{2}vm_{h}^{2}}{m^{2}}\eta\left(\xi m^{2},m_{h}^{2},p^{2}\right)\right.\nonumber \\
		&  & \left.+\frac{\left(d-1\right)e^{2}v}{m_{h}^{2}}\chi\left(m^{2}\right)+\frac{3e^{2}v}{2m^{2}}\chi\left(m_{h}^{2}\right)+\frac{e^{2}v}{2m^{2}}\chi\left(\xi m^{2}\right)\right]
		\end{eqnarray}
		The eleventh term, the one-loop correction shown in frame $\fbox{11}$ of FIG.~\ref{Y22}, gives
	\begin{eqnarray}
	ivp_{\nu}\langle \left(\rho\partial_{\mu}h\right)\left(p\right)\rho\left(-p\right)\rangle  & = & \frac{m_{h}^{2}}{2}\frac{1}{p^{2}+\xi m^{2}}\mathcal{L}_{\mu\nu}\left(p\right)\left[\left(m_{h}^{2}-\xi m^{2}-p^{2}\right)\eta\left(\xi m^{2},m_{h}^{2},p^{2}\right)\right.\nonumber \\
	&  & \left.+\chi\left(m_{h}^{2}\right)-\chi\left(\xi m^{2}\right)\right]
	\end{eqnarray}
			The twelfth term, the one-loop correction shown in frame $\fbox{12}$ of FIG.~\ref{Y22}, gives
	\begin{eqnarray}
	\frac{1}{4}p_{\mu}p_{\nu}\langle (hp)\left(p\right)(hp)\left(-p\right)\rangle  & = &\frac{ p^{2}}{4}\eta\left(\xi m^{2},m_{h}^{2},p^{2}\right)\mathcal{L}_{\mu\nu}\left(p\right)
	\end{eqnarray}
		The thirteenth term, the one-loop correction shown in frame $\fbox{13}$ of FIG.~\ref{Y22}, gives
	\begin{eqnarray*}
		-\langle \left(\rho\partial_{\mu}h\right)\left(p\right)\left(h\partial_{\nu}\rho\right)\left(-p\right)\rangle  & = & -\frac{1}{4\left(d-1\right)p^{2}}\mathcal{P}_{\mu\nu}\left(p\right)\left\{ \left[\left(-m_{h}^{2}+\xi m^{2}+p^{2}\right)^{2}+4m_{h}^{2}p^{2}\right]\eta\left(m_{h}^{2},\xi m^{2},p^{2}\right)\right.\\
		&  & \left.+\left(-m_{h}^{2}+\xi m^{2}-p^{2}\right)\chi\left(\xi m^{2}\right)-\left(-m_{h}^{2}+\xi m^{2}+p^{2}\right)\chi\left(m_{h}^{2}\right)\right\} \\
		&  & -\frac{1}{4p^{2}}\mathcal{L}_{\mu\nu}\left(p\right)\left\{ -\left(m_{h}^{2}-\xi m^{2}-p^{2}\right)\left(m_{h}^{2}-\xi m^{2}+p^{2}\right)\eta\left(\xi m^{2},m_{h}^{2},p^{2}\right)\right.\\
		&  & +\left(m_{h}^{2}-\xi m^{2}-p^{2}\right)\chi\left(\xi m^{2}\right)\\
		&  & \left.-\left(m_{h}^{2}-\xi m^{2}+p^{2}\right)\chi\left(m_{h}^{2}\right)\right\} 
	\end{eqnarray*}
		The fourteenth term, the one-loop correction shown in frame $\fbox{14}$ of FIG.~\ref{Y22}, gives
		\begin{eqnarray*}
			\frac{v^{2}}{4}p_{\mu}p_{\nu}\langle \rho\left(p\right)\rho\left(-p\right)\rangle_{1-{\rm loop}}  & = &  \frac{1}{\left(p^{2}+\xi m^{2}\right)^{2}}\frac{p^{2}}{4}\left\{ \left[m_{h}^{4}-\left(p^{2}+m_{h}^{2}\right)^{2}\right]\eta\left(\xi m^{2},m_{h}^{2},p^{2}\right)\right.\\
			&  & +\left[\left(m_{h}^{2}-m^{2}+p^{2}\right)^{2}+4m^{2}p^{2}\right]\eta\left(m^{2},m_{h}^{2},p^{2}\right)+p^{2}\chi\left(\xi m^{2}\right)\\
			&  & \left.+\left(m_{h}^{2}-m^{2}\right)\chi\left(m_{h}^{2}\right)-\left(m_{h}^{2}-m^{2}+p^{2}\right)\chi\left(m^{2}\right)\right\} \mathcal{L}_{\mu\nu}\left(p\right)
		\end{eqnarray*}

		\bibliographystyle{FMSU1_vdudal}
		
		\bibliography{reb}
		
	\end{document}